\documentclass[]{aa}

\usepackage{booktabs}
\usepackage{longtable}
\usepackage{graphicx}
\usepackage{threeparttablex} 

\usepackage{array}
\usepackage{changepage}
\usepackage{lineno}
\usepackage[flushmargin]{footmisc}
\usepackage{afterpage}

\usepackage{lscape}

\usepackage[]{xcolor}
\usepackage[normalem]{ulem}
\usepackage{hyperref}
\hypersetup{
    unicode=false,                  
    pdftoolbar=true,                
    pdfmenubar=true,                
    pdffitwindow=true,              
    pdfstartview={FitH},            
    pdftitle={S-PLUS USS},          
    pdfauthor={Perottoni},            
    pdfsubject={Astronomy},         
    pdfcreator={dvipdf},            
    pdfproducer={dvipdf},           
    pdfkeywords={}, 
    pdfnewwindow=true,              
    colorlinks=true,                
    linkcolor=red,                  
    citecolor=blue,                 
    filecolor=magenta,              
    urlcolor=cyan,                  
    breaklinks=true,
    linktocpage
}

\newcommand{\orcid}[1]{\href{https://orcid.org/#1}{\includegraphics[width=10pt]{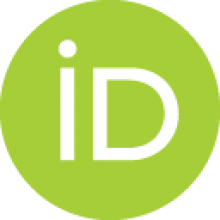}}}

\begin{document}

\title{The S-PLUS Ultra-Short Survey: first data release\thanks{All USS DR1 data is publicly available through the \href{https://splus.cloud}{https://splus.cloud} service.}}

\author{H\'elio D. Perottoni\inst{1,2}\orcid{0000-0002-0537-4146},
 Vinicius M.\ Placco\inst{3}\orcid{0000-0003-4479-1265},
 Felipe {Almeida-Fernandes}\inst{2}\orcid{0000-0002-8048-8717},
 F\'abio R. Herpich\inst{4}\orcid{0000-0001-7907-7884},
 Silvia Rossi\inst{2}\orcid{0000-0001-7479-5756},
  Timothy C. Beers\inst{5}\orcid{0000-0003-4573-6233},
  Rodolfo Smiljanic\inst{1}\orcid{0000-0003-0942-7855},
 Jo\~ao A. S. Amarante\inst{6,7}\orcid{0000-0002-7662-5475},
 Guilherme Limberg\inst{2}\orcid{0000-0002-9269-8287},
   Ariel Werle\inst{8}\orcid{0000-0002-4382-8081},
   Helio J. Rocha-Pinto\inst{9}\orcid{0000-0002-5274-4955},
   Leandro {Beraldo e Silva}\inst{10}\orcid{0000-0002-0740-1507},
   Simone Daflon\inst{11}\orcid{0000-0001-9205-2307},
   Alvaro {Alvarez-Candal}\inst{12,13}\orcid{0000-0002-5045-9675},
   Gustavo B {Oliveira Schwarz}\inst{2,14}\orcid{0009-0003-6609-1582},
 William Schoenell\inst{15}\orcid{0000-0002-4064-7234},
 Tiago Ribeiro\inst{16}\orcid{0000-0002-0138-1365},
 Antonio Kanaan\inst{17}\orcid{0000-0002-2484-7551}}

\titlerunning{The S-PLUS Ultra Short Survey}
\authorrunning{Perottoni et al. 2023}

\institute{Nicolaus Copernicus Astronomical Center, Polish Academy of Sciences, ul. Bartycka 18, 00-716, Warsaw, Poland\\ 
              \email{hperottoni@gmail.com}
\and
Universidade de S\~ao Paulo, Instituto de Astronomia, Geof\'isica e Ci\^encias Atmosf\'ericas, Departamento de Astronomia, SP 05508-090, S\~ao Paulo, Brasil
\and
NSF NOIRLab, Tucson, AZ 85719, USA
\and
Institute of Astronomy, University of Cambridge, Madingley Road, Cambridge, CB3 0HA, UK
\and
Department of Physics and Astronomy and JINA Center for the Evolution of the Elements, University of Notre Dame, Notre Dame, IN 46556, USA
\and
Institut de Ci\'encies del Cosmos (ICCUB), Universitat de Barcelona (UB), Mart\'i i Franqu\'es, 1, 08028 Barcelona, Spain
\and{Visiting Fellow, Jeremiah Horrocks Institute, University of Central Lancashire, Preston, PR1 2HE, UK}
\and
INAF-Osservatorio Astronomico di Padova, Vicolo dell’Osservatorio 5, 35122 Padova, Italy
\and
Universidade Federal do Rio de Janeiro, Observat\'orio do Valongo, Lad. Pedro Ant\^onio 43, 20080-090, Rio de Janeiro, Brazil
\and
Department of Astronomy \& Steward Observatory, University of Arizona, Tucson, AZ, 85721, USA
\and
Observatório Nacional, MCTI, Rua Gal. José Cristino 77, Rio de Janeiro, 20921-400, RJ, Brazil
\and
Instituto de Astrof\'isica de Andaluc\'ia, CSIC, Apt 3004, E18080 Granada, Spain
\and
Instituto de F\'isica Aplicada a las Ciencias y las Tecnolog\'ias, Universidad de Alicante, San Vicent del Raspeig, E03080, Alicante, Spain\
\and
Universidade Presbiteriana Mackenzie, Rua da Consolac\~ao, 930, S\~ao Paulo, 01302-907, SP, Brazil
\and
GMTO Corporation 465 N. Halstead Street, Suite 250 Pasadena, CA 91107, USA
\and
Rubin Observatory Project Office, 950 N. Cherry Ave., Tucson, AZ 85719, USA
\and
Departamento de F\'isica, Universidade Federal de Santa Catarina, Florian\'opolis, SC 88040-900, Brazil
}

\abstract{This paper presents the first public data release of the S-PLUS Ultra-Short Survey (USS), a photometric survey with short exposure times, covering approximately 9300 deg$^{2}$ of the Southern sky. The USS utilizes the Javalambre 12-band magnitude system, including narrow and medium-band and broad-band filters targeting prominent stellar spectral features. The primary objective of the USS is to identify bright, extremely metal-poor (EMP; [Fe/H] $\leq -3$) and ultra metal-poor (UMP; [Fe/H] $\leq -4$) stars for further analysis using medium- and high-resolution spectroscopy.}{This paper provides an overview of the survey observations, calibration method, data quality, and data products. Additionally, it presents the selection of EMP and UMP candidates.}{The data from the USS were reduced and calibrated using the same methods as presented in the S-PLUS DR2. An additional step was introduced, accounting for the offset between the observed magnitudes off the USS and the predicted magnitudes from the very low-resolution Gaia XP spectra.}{This first release contains data for 163 observed fields totaling $\sim$324 deg$^{2}$ along the Celestial Equator. The magnitudes obtained from the USS are well-calibrated, showing a difference of $\sim 15$ mmag compared to the predicted magnitudes by the GaiaXPy toolkit. By combining colors and magnitudes, 140 candidates for EMP or UMP have been identified for follow-up studies.}{The S-PLUS USS DR1 is an important milestone in the search for bright metal-poor stars, with magnitudes in the range 10 $ < r \leq 14$. 
The USS is an ongoing survey; in the near future, it will provide many more bright metal-poor candidate stars for spectroscopic follow-up.} 

\keywords{Surveys -- Techniques: photometric -- Methods: data analysis -- stars: metal-poor -- Galaxy: evolution -- Galaxy: formation}

\maketitle


\section{Introduction} \label{sec:intro}

The early Universe was characterized by the lack of heavy elements that were gradually formed through the various nucleosynthetic burning and explosive stages of stellar evolution by different nuclear processes \citep[e.g.,][]{Burbidge1957,Cameron1957}. Its chemical and enrichment history can be better understood by studying the oldest stars with lifetimes sufficiently long to still be found in the Milky Way today. These ancient stars hold in their atmospheres valuable information about the chemical composition of the early Universe and how it evolved to its present-day makeup \citep{BeersChristlieb2005,FrebelNorris2015}. In this context, metal-poor stars play a fundamental role in the study of (i) astrophysical sites for the formation of different elements \citep{Nomoto2013}, (ii) the metal-mixing processes that affect the formation of subsequent stellar generations \citep[e.g.,][]{YangKrumholz2012mix, PetitKrumholz2015mix}, (iii) the earliest chemical evolution at high redshift ($z > 20$) \citep{Bromm2004, Bromm2011review}, and (iv) the assembly history of the Milky Way \citep{Yuan2020dtgs, Limberg2021dtgs,Gudin2021,Shank2022dtgs,daSilva&Smiljanic2023,Zepeda2023}.

To address these questions, significant efforts have been devoted over the past 50 years to systematically identify very metal-poor (VMP; [Fe/H]\footnote{Definition of the abundance of a star ($\star$) relative to the Sun ($\odot$): [A/B] $= \log (\rm N_A/N_B)_\star - \log (\rm N_A/N_B)_\odot$, where $\rm N_A$ ($\rm N_B$) is the number of atoms of element A (B).} 
$\leq -2$), extremely metal-poor (EMP; [Fe/H] $\leq -3$), and ultra metal-poor (UMP; [Fe/H] $\leq -4$) stars \citep{BeersChristlieb2005}. Early efforts relied on photographic objective-prism surveys followed up with medium-resolution spectroscopy, such as the HK survey \citep{beers1985,beers1992} and the Hamburg/ESO survey \citep{Christlieb2003,Christlieb2008}. Later spectroscopic surveys that provided significant numbers of metal-poor stars include the Sloan Digital Sky Survey (SDSS; \citealt{sdssYork}), and its sub-surveys, the Sloan Extension for Galactic Understanding and Evolution (SEGUE; 
\citealt{Yanny2009, Rockosi2022}) and the Apache Point Observatory Galactic Evolution Experiment (APOGEE; \citealt{Majewski2017}), 
the Radial Velocity Experiment (RAVE; \citealt{RAVE, placco2018rave}), the Large Sky Area Multi-object Fiber Spectroscopic Telescope (LAMOST; \citealt{LAMOST2012,LI2018}), and the GALactic Archaeology with HERMES (GALAH; \citealt{deSilva2015}).

The numbers of recognized metal-poor stars has recently been boosted (\citealt{Xu2022b,Andrae2023,Lu2023,Yao2023}) by the use of the Gaia DR3 XP Spectra (\citealt{GaiaCollaboration2016,GaiaCollaboration2023_dr3,DeAngeli2023}) and by the use of multi-band photometric surveys, such as the Best \& Brightest survey (B\&B; \citealt{BeB, Placco2019beb, Limberg2021beb, Xu2022}). The use of narrow and medium-band photometry of metallicity-sensitive features such as the SkyMapper Southern Sky Survey (SMSS; \citealt{SkyMapper,SkyMapperDR2, Chiti2021smss,Huang2022}), the Pristine survey (\citealt{Pristine1,martin2023}), the Stellar Abundances and Galactic Evolution Survey (SAGES; \citealt{Zheng2018,Fan2023,Huang2023}), the Javalambre Photometric Local Universe Survey (J-PLUS; \citealt{JPLUSsurvey}), and 
the Southern Photometric Local Universe Survey (S-PLUS; \citealt{SPLUSsurvey}), have led to an increased success in identifying new metal-poor stars (\citealt{Dacosta2019,Galarza2022,Placco2022,Yang2022,Almeida-Fernandes2023,Placco2023}; Huang et al., in prep.).

The increase in the number of metal-poor stars combined with Gaia data has been crucial to revolutionizing our view of the Milky Way's structure and evolution through Galactic Archaeology \citep[e.g.,][]{Helmi2020}. In the Galactic halo, a handful of accretion events have been revealed \citep{koppelman2019, naidu2020, malhan2022} as well as various tidal streams \citep{malhan2018streams, ibata2019streams, yuan2020, ibata2021streams, dodd2023, Ibata2023}. Moreover, the Galactic disk system also exhibits a very and extremely metal-poor component of unknown (but possibly primordial) origin \citep{Sestito2019,Sestito2020,DiMatteo2020,Carter2021,Cordoni2021,Hong2023,Zhang2023}; see simulation efforts by \citet{Sestito2021}, \citet{Santistevan2021}, \citet{Hirai2022}, and \citet{Sotillo-Ramos2023}. Furthermore, an old, metal-poor stellar population corresponding to remnants of the earliest phase of galaxy formation has been identified in the inner parts of the Galaxy \citep{Kruijssen2019,Horta2021,BelokurovKravtsov2022,rix2022,XiangRix2022}, including the discovery of new EMP stars in the bulge region \citep{Howes2015,Howes2016,Arentsen2020,Reggiani2020}.

In particular, EMP and UMP stars are likely to have originated from pristine gas that was relatively free of heavier elements, providing valuable insights into the initial chemical enrichment and the properties of the first massive Population III stars and their subsequent supernovae \citep[e.g.,][]{Bromm2004, Iwamoto2005,Nomoto2013,FrebelNorris2015,Jeon2021,Koutsouridou2023}. 
Consequently, the chemical abundance patterns of EMP and UMP stars can place direct constraints on the nature of the first stars formed in the Universe; see also \citet{Hartwig2018,Hartwig2019,Hartwig2023} and \citet{Hansen2020}. Additionally, it has also been shown that more than 80\% of the observed UMP stars in the Galaxy are carbon enhanced \citep[e.g.,][]{Lee2013,Placco2014,Yoon2018}. These carbon-enhanced metal-poor stars (CEMP; [C/Fe] $> +0.7$ and [Fe/H] $< -1.0$, see \citealt{BeersChristlieb2005} and \citealt{Aoki2007}) have sparked a great deal of recent interest. 

The EMP and UMP stars are rare \citep{FrebelNorris2013book}. The majority found to date are relatively faint, and present very weak spectral lines, making their detailed abundance analyses through high-spectral resolution observations challenging, time-consuming, and sometimes not even feasible for some wavelength ranges, even with 8-10 meter class telescopes. Therefore, it is highly desirable to identify relatively bright EMP and UMP and CEMP stars in order to obtain their detailed spectra, also including the near-ultraviolet spectral region \citep{Ernandes2023,Bonifacio2023}, increasing the amount of abundance information that can be obtained, \citep[e.g.,][]{Placco2014b, Placco2015b,Shejeelammal2024}, and thereby better constrain theoretical nucleosynthesis models as well as the initial mass function of the first stars \citep{Umeda2005,heger2010,Meynet2010,Nomoto2013,Jeon2021,Koutsouridou2023}.

To address the present dearth of relatively bright EMP and UMP stars, we have developed the S-PLUS Ultra-Short Survey (USS) as a sub-survey of S-PLUS. The USS utilizes exposures that are approximately 1/36th of the nominal values used in the S-PLUS survey \citep{SPLUSsurvey}, covering the same 12 photometric bands, and thereby enable the identification of relatively bright VMP and EMP stars \citep{Whitten2019,whitten2021}. By extending the brightness limit by at least four magnitudes, the USS enables observations of brighter sources compared to other narrow-band surveys by avoiding the saturation limitations of previous surveys. Additionally, the identification of relatively bright EMP and UMP stars opens up the possibility of conducting spectroscopic studies in the near-ultraviolet regime using the Hubble Space Telescope \citep[e.g.,][]{Placco2014,Placco2015a,Roederer2016,Holmbeck2020,Roederer2022}, while it is still available.


This paper is organized as follows.  Section~\ref{sec:data} outlines the USS design, implementation, and current observing completion status. Section~\ref{sec:calib} describes the data reduction and calibration of the USS data, followed by details on Data Release 1 (DR1) in Section~\ref{sec:photometry}. Preliminary tests on the search for very low-metallicity stars in the USS DR1 are provided in Section~\ref{sec:search}.  Section~\ref{sec:conclusion} presents our conclusions and perspectives for future work.

\section{The S-PLUS Ultra-Short Survey} 
\label{sec:data}

The S-PLUS project, facilitated by the S-PLUS Consortium, was built to be the Southern Hemisphere counterpart of J-PLUS, and is operated in collaboration with the Astronomy Department at the University of São Paulo, Brazil. The S-PLUS survey includes various sub-surveys, one of which is known as the S-PLUS Ultra-Short Survey (USS). The USS is an imaging survey that covers the same area as the overall S-PLUS Main Survey, but with much shorter exposure times. As in the S-PLUS Main survey, USS employs a set of 12 bands (seven narrow and medium-band filters and five broad-band filters) for its observations, enabling comprehensive characterization of objects by imaging them in different regions of the optical spectra, with the narrow and medium-band filters placed on crucial elemental absorption lines. This section provides a concise overview of the instruments and observations involved in the USS.

\subsection{Survey Instrumentation}
\label{subsec:surveys}


The USS is an imaging survey conducted at the Cerro Tololo Inter-American Observatory in Chile, situated at an altitude of approximately 2200 meters. It employs the same telescope, camera, and filters as the S-PLUS to ensure consistency across the projects.

The dedicated telescope for the USS is called the T80-South (T80S), a 0.826m robotic telescope specifically designed for wide-field optical imaging. Although the telescope is automated using the observatory control system \verb/chimera/\footnote{\url{https://github.com/astroufsc/chimera/}}, the observing staff has access to tools that display cloud cover, weather conditions, and observing status to monitor the progress of the observations. The camera used in the USS is also controlled by the {\tt chimera} system. This camera is designed to capture wide-field images with dimensions of 1.4 by 1.4 degrees, using a CCD of 9232 $\times$ 9216 pixels (see \citealt{MarinFranch2012camera}) and a plate scale of 0.55 arcsec pixel$^{-1}$. The detector is read out via 16 amplifiers arranged in an array of eight columns and two rows, following the same mode as the S-PLUS Main Survey (see Table 1 of \citealt{SPLUSsurvey}).

The USS adopts the Javalambre filter system \citep{JPLUSsurvey}, the same 12 bands that the S-PLUS Main Survey uses. Of those, seven are narrow and medium-band filters designed to map stellar spectral features, and five are SDSS-like broad-bands $ugriz$ filters (see Table \ref{tab:shortexposure}), allowing for the estimation of stellar parameters such as effective temperature ($T_{\rm eff}$), surface gravity ($\log g$), and metallicity ([Fe/H]) \citep{gruel2012}, as well as elemental-abundance estimates for a limited additional number of species (e.g., C, N, Mg, and Ca).  Huang et al. (in prep.) provide calibrations that are used for the determination of C and Mg abundance estimates; N and Ca will be added in the near future.
For more technical details on the survey instrumentation, see \citet{SPLUSsurvey}.

\subsection{Survey Design}
\label{subsec:surveys}

\begin{figure*}[ht!]
\centering
\includegraphics[width=\textwidth]{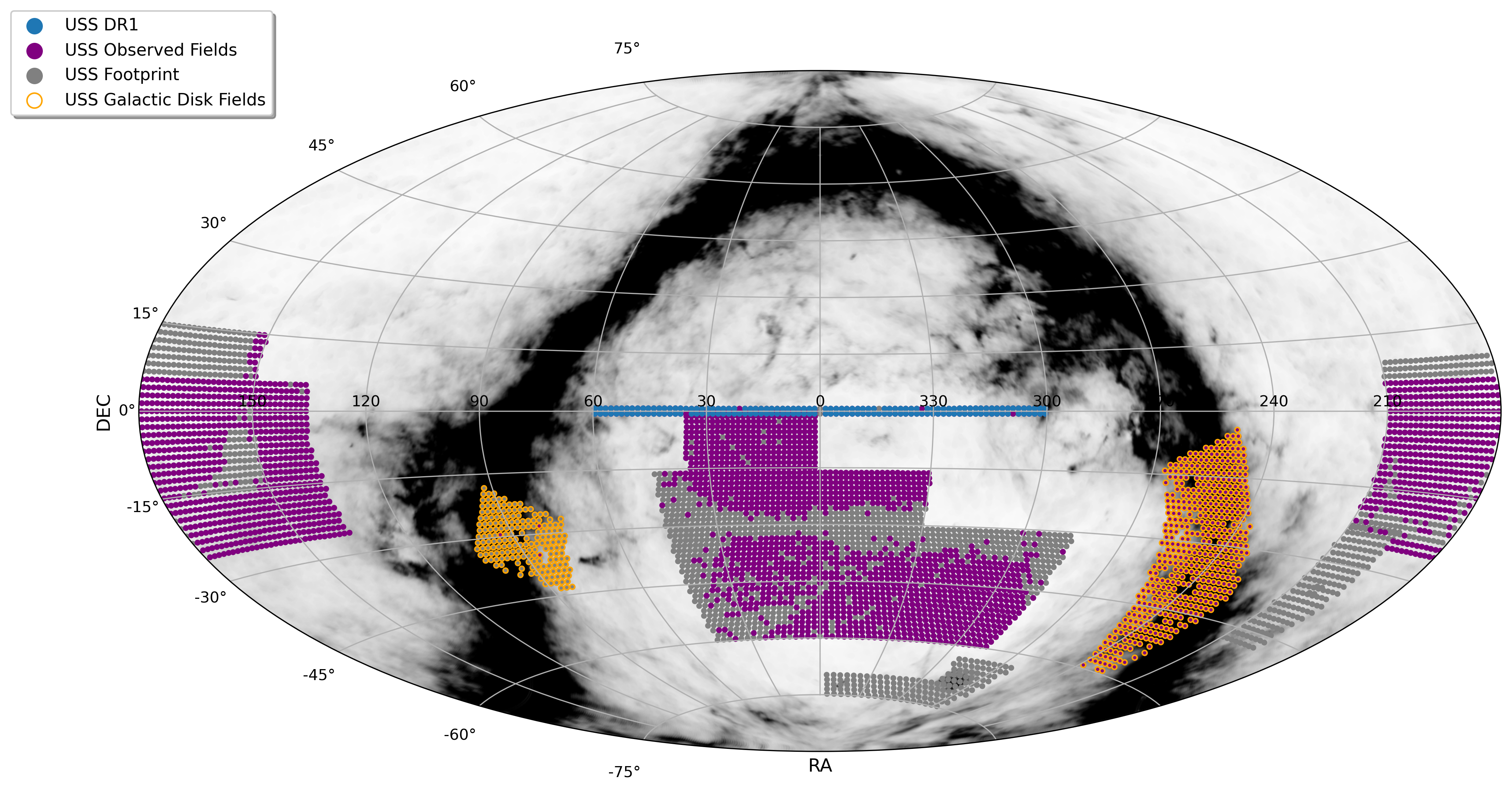}
\caption{USS footprint in equatorial coordinates using an Aitoff projection. The blue circles are the USS DR1 fields. The purple circles are the observed fields, and the gold ring indicates the fields in the region of the Galactic disk. The gray circles indicate the footprint area not observed yet. The background represents E$(B-V)$ values ranging from 0.0 (white) to 0.5 (black) according to the map by \citet{schlegel1998}.  }
\label{fig:splusfootprint}
\end{figure*}


The USS will cover 9300 deg$^{2}$ of the Southern Hemisphere sky when completed. The survey footprint is shown in Figure \ref{fig:splusfootprint} and follows exactly the same area as the S-PLUS Main Survey. The USS can be separated into two regions: the fields within the S-PLUS Galactic sub-survey area (represented by gold rings in Figure \ref{fig:splusfootprint}) and regions avoiding the Galactic plane (gray and purple symbols). This distinction is necessary because the region near the Galactic plane is significantly more crowded and requires point spread function (PSF) photometry. However, including these fields is not the focus of this paper; rather, it is a topic for future data releases.

The area included in DR1 is divided into 163 tiles (blue circles) along the Celestial Equator, each measuring 2 deg$^{2}$. USS operations began in October 2018 and are expected to last seven years. At the time of this writing, data has been acquired for 3332 tiles in all the 12 bands (purple circles), which accounts for $\sim 70\%$ completion of the USS observations. The coverage area of the observed fields is much larger than that provided in the USS DR1, which serves as a limited sample used to test and refine the method before applying it to all observed fields.

Each tile was observed in a single epoch and had only one exposure per filter, with a total integration time of 110 seconds per tile. The exposure times of the USS (see Table \ref{tab:shortexposure}) are approximately $1/36$th of those from the S-PLUS Main Survey for each filter. Consequently, the saturation limit of the USS is $\sim10$ mag for the $r$-band, or about four magnitudes brighter than for the S-PLUS Main Survey.

\begin{table}
\tabcolsep 0.08truecm
\centering
\caption{USS filter summary and exposure times.}
\label{tab:shortexposure}
\begin{tabular}{cccc|cccc} 
\hline
Filter  &  $\lambda_{\rm eff}$ & Feature & T$_{\rm exp}$ &  Filter & $\lambda_{\rm eff}$ & Feature & T$_{\rm exp}$    \\
name    &  [\AA]& &     (s)     &  name     & [\AA]& & (s)        \\
\hline
\hline
$u$     &  3533&      & 20      &  J0515    & 5133&   Mgb Triplet      &5          \\
J0378   &  3773&  $[\mathrm{O}\,\textsc{ii}]$, CN    &19      &  $r$      & 6251&        &3          \\
J0395   &  3940&  Ca H+K &10      &  J0660    & 6613&  H$\alpha$      &24         \\
J0410   &  4095&  H$\delta$    &5       &  $i$      & 7670&        &4          \\
J0430   &  4292&  CH G-band     &5       &  J0861    & 8607&   Ca Triplet     &7          \\
$g$     &  4758&      &3       &  $z$      & 8936&        &5          \\
\hline
\hline
\end{tabular}
\end{table}

Given that the USS was developed to optimize usage of the telescope, the observations were taken during all levels of Moon brightness, with a minimum distance of 40\,deg from the Moon. In the USS DR1 the median airmass of the observed fields is $\sim$ 1.28, with a standard deviation of 0.14 (see Table \ref{tab:airmass}). The current physical limit of the observations is set at an altitude of 35\,deg, corresponding to an airmass of $\sim 1.7$. The fields of the USS that constitute this study point towards regions with low--extinction and present an $\langle E_{(B-V)} \rangle$ = 0.068 mag.

\begin{table}
\tabcolsep 0.2truecm
\centering
\caption{USS filters mean and standard deviations of the airmass and FWHM spatial coverage.}
\label{tab:airmass}
\begin{tabular}{ccccc} 
\hline
Filter  &  Airmass & Airmass &  FWHM &  FWHM     \\
        &          &         &  (arcsec) & (arcsec)      \\
        &  Mean    & Std     & Mean  &  Std     \\
\hline
\hline
$u$     &  1.28 & 0.13 & 1.98 & 0.60      \\
J0378   &  1.28 & 0.13 & 1.86 & 0.56      \\
J0395   &  1.28 & 0.13 & 1.80 & 0.57      \\
J0410   &  1.28 & 0.13 & 1.77 & 0.57      \\
J0430   &  1.29 & 0.14 & 1.75 & 0.57      \\
$g$     &  1.29 & 0.14 & 1.72 & 0.60      \\
J0515    & 1.29 & 0.14 & 1.68 & 0.65       \\
$r$      & 1.29 & 0.14 & 1.42 & 0.53       \\
J0660    & 1.29 & 0.15 & 1.69 & 0.58       \\
$i$      & 1.29 & 0.14 & 1.50 & 0.53       \\
J0861    & 1.29 & 0.15 & 1.67 & 0.65       \\ 
$z$      & 1.29 & 0.15 & 1.41 & 0.52       \\
\hline
\hline
\end{tabular}
\end{table}

\section{Calibration Method}
\label{sec:calib}

The observed images are processed and calibrated similarly to the observations for the S-PLUS Main Survey. Here, we briefly describe this process and the unique procedures used for the USS data.

\subsection{Data Reduction and Astrometry}
\label{subsec:reduction}

The reduction, comprising overscan and bias subtraction, flat-field correction, and fringe subtraction, uses the same processes as for the Main Survey described in \citet{SPLUSsurvey}. The tool deployed is the \texttt{jype} pipeline version 0.9.9, the same as used for the S-PLUS data releases to date. The only step of the processing we skip is the co-adding, given that the USS consists of only one observation per filter.

Astrometry is calculated using the SCAMP code \citep{scamp} and the point source catalog of the Two Micron All Sky Survey (2MASS; \citealt{Skrutskie2006}) as a reference. We obtain a dispersion of $\sim 135$\,mmas for both RA and DEC, similar to S-PLUS DR2 \citep{Almeida-Fernandes+2022}. It is worth noting that being a shallow survey, individual stars can have different sky coordinates compared with other surveys depending on the epoch of the observations due to high proper motion.

\subsection{Photometry Extraction}

The USS DR1 provides circular aperture photometry obtained using SExtractor \citep{sextractor}. The final catalogs contain magnitudes measured in 3- and 6-arcsec diameter apertures (labeled APER\_3 and APER\_6) and the aperture-corrected instrumental magnitudes (PStotal). The PStotal apertures correspond to the 3-arcsec apertures, corrected by the amount of flux that the source emits outside this aperture. This correction is obtained by measuring the magnitudes in 32 concentric apertures centered around each source, and computing the average change in magnitude in increasingly larger apertures until the changes converge to zero. Only sources with a signal-to-noise ratio (S/N) between 30 and 1000 and a CLASS\_STAR parameter with a value greater than 0.9 are considered in this step. This correction is computed for every observation in each filter individually, and depends mostly on the atmospheric conditions at the time of the observation (which dictates the seeing and, consequently, the PSF). The PStotal magnitude corresponds to the total magnitude of the source, provided it behaves as a point source in the observation.

\subsection{Photometric Calibration Method}
\label{subsec:astrometry}

In this section, we discuss the techniques applied to perform the photometric calibration of the USS DR1. The PStotal magnitudes used in this section are first corrected for interstellar medium (ISM) extinction using the maps from \citet{schlegel1998}. The magnitudes in the reference catalog are also corrected for ISM extinction in the same way. The extinction coefficients for the S-PLUS filters, which are necessary to correct the ISM extinction, are different from the J-PLUS ones (the coefficients are a function of the transmission curve, which in turn are obtained from the instrument response and atmospheric conditions of the site). To obtain the coefficients, we have used the extinction curve obtained from \citet{Schlafly2016}\footnote{\url{https://e.schlaf.ly/apored/extcurve.html}}, with R$_v$= 3.1. The transmission curves and the reference wavelengths are obtained from the Spanish Virtual Observatory filter profile service\footnote{\url{http://svo2.cab.inta-csic.es/theory/fps/index.php?mode=browse&gname=CTIO&gname2=S-PLUS&asttype=}}; the coefficients for each of the 12 filters are given in Table~\ref{tab:extinction}.

\begin{table}
\tabcolsep 0.4truecm
\centering
\caption{Extinction coefficients for the S-PLUS filters.}
\label{tab:extinction}
\begin{tabular}{cccc} 
\hline
Filter  &  $\lambda_{ref}$ & k$_{\lambda}$ &  A$_{\lambda}$/A$_{V}$     \\
        & [\AA]    &      &       \\
\hline
\hline
$u$     &  3533.29 & 4.937 & 1.593       \\
J0378   &  3773.13 & 4.664 & 1.505       \\
J0395   &  3940.70 & 4.480 & 1.445       \\
J0410   &  4095.27 & 4.316 & 1.392       \\
J0430   &  4292.39 & 4.113 & 1.327       \\
$g$     &  4758.49 & 3.663 & 1.182       \\
J0515    & 5133.15 & 3.334 & 1.075       \\
$r$      & 6251.83 & 2.515 & 0.811       \\
J0660    & 6613.88 & 2.304 & 0.743       \\
$i$      & 7670.59 & 1.803 & 0.582       \\
J0861    & 8607.59 & 1.458 & 0.470       \\
$z$      & 8936.64 & 1.416 & 0.457       \\
\hline
\hline
\end{tabular}
\end{table}

The photometric calibration technique applied in the USS takes advantage of the large field of view of the S-PLUS instrument. Each observation contains hundreds of stars with precise photometry and well-known magnitudes in the literature. In this work, for the photometric calibration, we used the reference magnitudes from the ATLAS All-Sky Stellar Reference Catalog \citep[ATLAS Refcat2;][]{Tonry+2018}. This catalog provides an almost all-sky coverage with accurate and precise magnitudes for approximately one billion stars down to the AB-magnitude $\sim 19$, achieved by compiling the photometry from several surveys, including 2MASS, SMSS, Pan-STARRS DR1 \citep{Chambers+2016,Flewelling+2020}, and Gaia DR2 \citep{GaiaCollaboration2016, GaiaCollaboration2018}.

The ATLAS Refcat2 provides magnitudes on the Pan-STARRS magnitude system \citep{Chambers+2016}. These filters are similar, but not identical, to the filters $g$, $r$, $i$, and $z$ in S-PLUS. In addition, the reference catalog does not contain filters similar to the narrow bands. To extract reference magnitudes in the S-PLUS filter system from the data in the reference catalog, we employ the technique presented in \citet{Almeida-Fernandes+2022}. In summary, this is done by fitting synthetic stellar spectral energy distributions (SEDs) to the reference magnitudes using chi-square minimization. The synthetic SEDs library is obtained by convolving the synthetic spectra from \citet{Coelho2014} with the transmission curves of the reference catalog and the S-PLUS filters.

In practice, the SED fitting process allows us to convert the reference magnitudes in the Pan-STARRS system to reference magnitudes in the S-PLUS system. The calibration zero-points are then obtained by computing the difference between reference and instrumental magnitudes for dozens of stars in each observation. The zero-points are characterized as the mode of the distribution, and since they are estimated using the stars in the science images, they already consider the effects of sky transparency, airmass, and the instrument's sensitivity for that particular observation.

In the case of the USS, the use of the Refcat2 magnitudes limits the SED-fitting-based calibration to the seven reddest filters, as there is no coverage in the blue range of the spectrum to constrain the models and estimate reliable reference magnitudes for these filters. Therefore, the $u$, $J0378$, $J0395$, $J0410$, and $J0430$ filters are calibrated using a stellar locus-based technique, also presented in \citet{Almeida-Fernandes+2022}. In this technique, the stellar locus of the observation in the $y-g \times g-i$ space, where $y$ represents a blue filter (with $g$ and $i$ already previously calibrated), is compared to a reference stellar locus. The difference between the observation and the reference stellar locus is then used to characterize the calibration zero-point for the $y$-band.

After applying the SED-fitting-based calibration for the seven redder filters and the subsequent stellar-locus-based calibration for the five bluer filters, we performed an additional run of the SED-fitting-based calibration. This time, the chi-square minimization is done using the 12 pre-calibrated S-PLUS magnitudes. The use of the 12 filters contributes to better constraining of the synthetic SED models, and was found by \citet{Almeida-Fernandes+2022} to improve the calibration by providing corrections on the order of 10 mmag.

\begin{figure}
\centering
\includegraphics[width=\columnwidth]{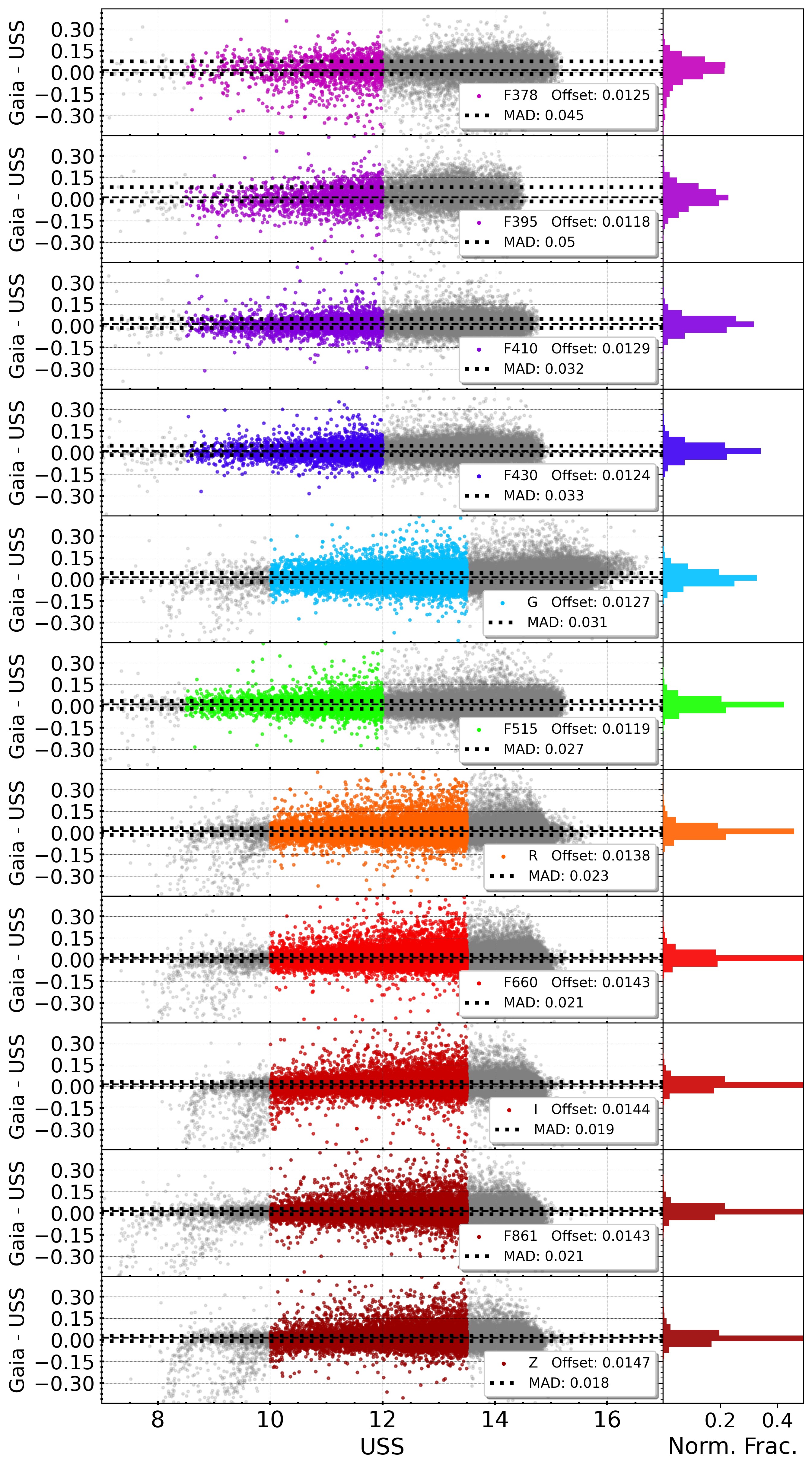}
\caption{Residual magnitude differences between S-PLUS magnitudes and the predicted S-PLUS XP spectra magnitudes are presented in each panel, as a function of the S-PLUS magnitudes. The corresponding offset for each filter is indicated in the upper right corner of the respective panel. The colored dots represent the region utilized for offset calculations. The gray dots are the sources with errors smaller than 0.04 mag. The black dotted lines indicate the Mean Absolute Deviation. In the right panels, the normalized fraction of the residual magnitude differences is shown.}
\label{fig:uss_gaia_predict}
\end{figure}

\subsection{Photometric Quality Assurance}
\label{subsec:astrometry}

To estimate the accuracy and precision of the photometric calibration, we compared the S-PLUS calibrated magnitudes to the convolved magnitudes obtained from  Gaia BP/RP spectra (hereafter XP spectra, \citealt{Montegriffo2023}). We obtained the S-PLUS photometry by synthesizing the flux in each S-PLUS filter from the Gaia XP photometry with the GaiaXPy\footnote{\url{https://gaia-dpci.github.io/GaiaXPy-website/}} python library.

The S-PLUS $u$-band (319.54--384.89 nm) extends beyond the blue edge of the wavelength range covered by Gaia BP/RP spectra of approximately 330 nm. Artificially truncating the $u$-band at 330 nm may result in significant differences between the observed and predicted magnitudes. To prevent this, we exclusively used eleven S-PLUS filters to compare the observed magnitudes with the predicted S-PLUS XP spectra magnitudes. We will examine comparisons of the calibration with numerical extrapolation techniques, as presented by \citet{Xiao2023jplus} for extrapolating the Gaia XP spectra, in a forthcoming data release.

In addition to the calibration method proposed by \citet{Almeida-Fernandes+2022}, we introduced a step that considers the offsets between S-PLUS magnitudes and the predicted S-PLUS XP spectra magnitudes. These offsets were then added to the internal step of the photometric calibration. To determine the offset for each filter, we selected objects within a magnitude range to avoid saturation and large photometric errors (< 0.02 mag). For the filters $J0378$, $J0395$, $J0410$, $J0430$, and $J0515$, we selected objects with magnitudes between 8.5 and 12, while for $g$, $r$, $J0660$, $i$, $J0861$, and $z$, objects within the magnitude range of 10 to 13.5 were considered. This difference arises because the redder filters appear to initiate saturation around 10th mag. As a conservative measure, we are adopting this as the limit regime; the classification of saturated objects will be explored in future data releases.
 
Figure \ref{fig:uss_gaia_predict} shows the final calibration accuracy in this work, represented by the residual magnitudes between the S-PLUS magnitudes and the predicted S-PLUS XP spectra magnitudes. The median offset for all filters is 15 mmag, and the average Mean Absolute Deviation (MAD) is 0.026 mag computed using data from the 163 fields in the USS DR1.

\section{Photometry Data}
\label{sec:photometry}

This section presents an overview of the data quality in USS DR1. We examine the zero-point distribution, detection completeness to all bands relative to the Gaia DR3 $G$-band, and the imaging depth achieved at various S/N levels. 

\begin{figure}
\centering
\includegraphics[width=\columnwidth]{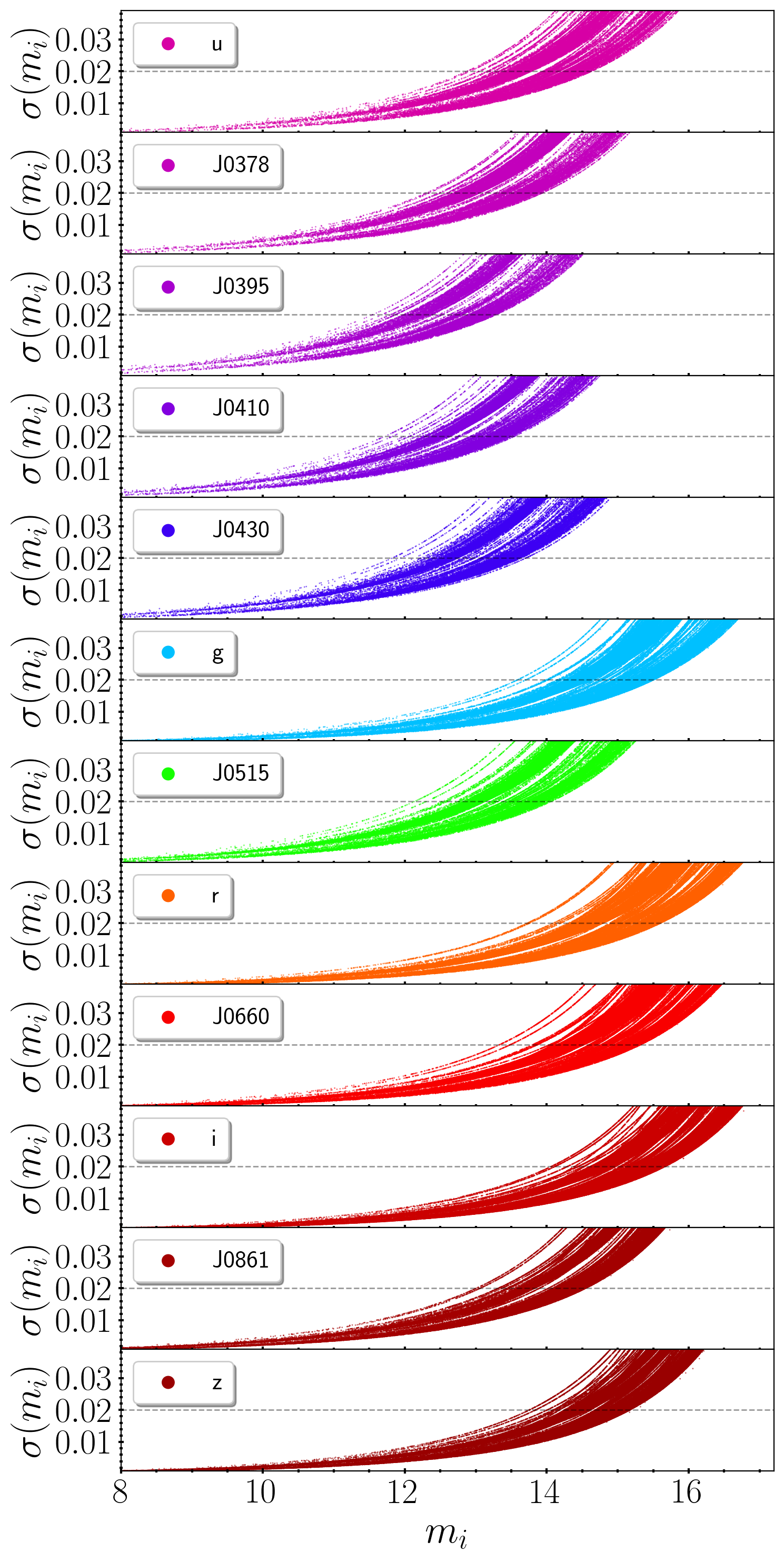}
\caption{Photometric error ($\sigma(m_{i})$) for each filter ($m_{i}$) of the USS. For visual reference, the gray-dashed line indicates where $\sigma(m_{i})$ = 0.02 mag. The corresponding filter is indicated in the upper left corner of the panel.}
\label{fig:mag_error}
\end{figure}

\begin{figure}
\centering
\includegraphics[width=\columnwidth]{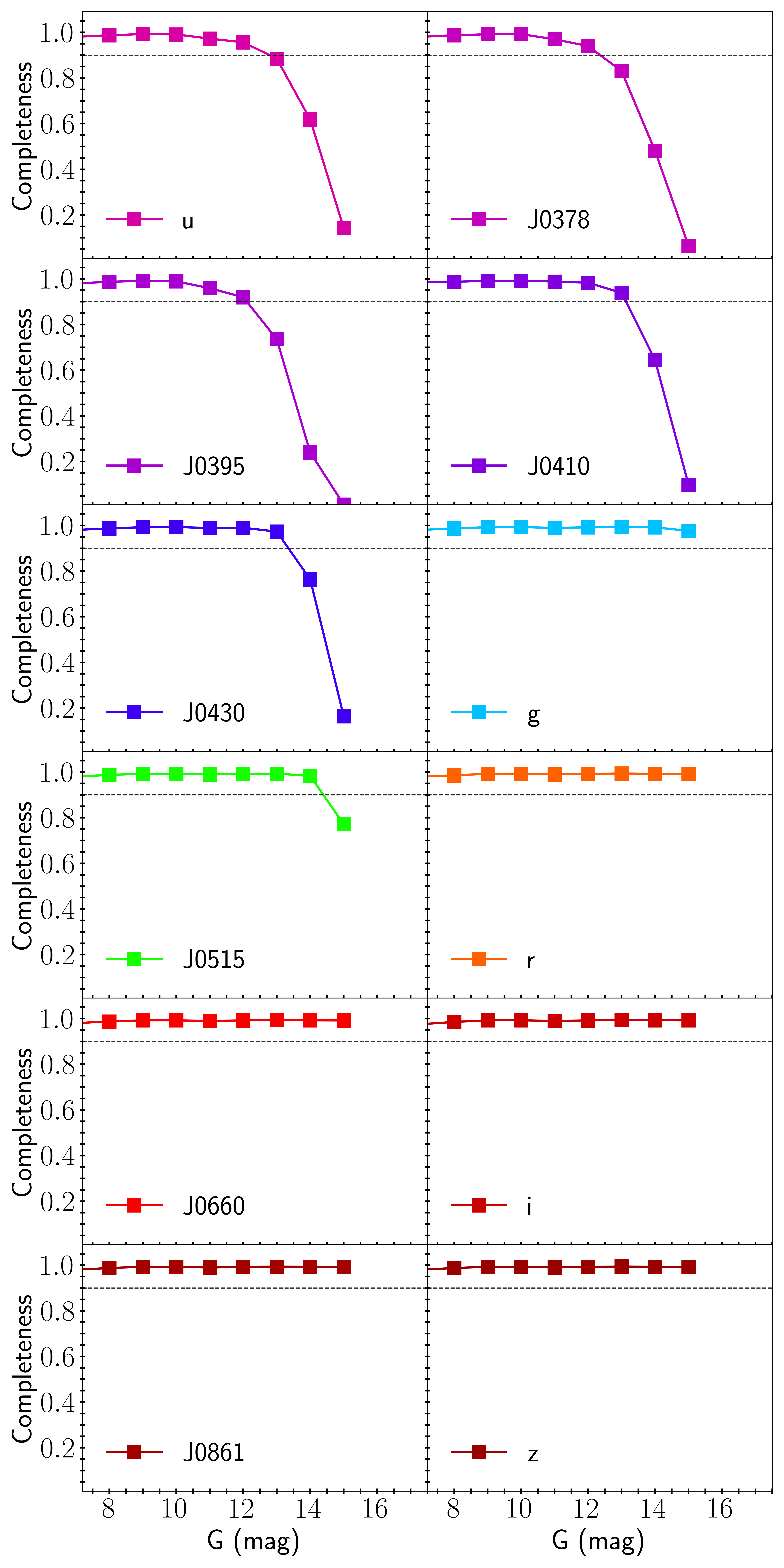}
\caption{Detection completeness in relation to the Gaia DR3 $G$-band for all filters from USS DR1. An internal legend indicates the filters that correspond to each panel. For visual reference, the gray dashed line indicates 90\% of sources detected. We considered the detection of all the objects with S/N > 3 if the SExtractor attributed a magnitude value other than 99 for this filter.}
\label{fig:uss_completeness}
\end{figure}

\subsection{Photometric Magnitude Error}

Photometric errors are influenced by factors such as exposure time, weather conditions, and instrument variations over time. Given that photometric conditions in the USS observations are not always ideal, with nights often not meeting the criteria of a dark night (e.g., observations in any moon phase) and good sky transparency (e.g., no clouds or low humidity), a larger dispersion in photometric errors for the same magnitude is expected. This dispersion is further amplified at faint magnitudes due to the smaller number of photons from the source.


Figure \ref{fig:mag_error} shows the photometric magnitude error for USS DR1 data. For bright sources, the photometric errors are negligible over a wide range of magnitudes. The dispersion in the magnitude error data arises primarily from the acquisition under different sky conditions.

Notably, the $g$, $r$, and $i$ filters exhibit higher S/N per magnitude, while the filters J0378, J0395, J0410, and J0430 display lower S/N, which can be attributed to their respective exposure times. These relationships vary slightly from those in the Main Survey due to the short exposure times, which are only a few seconds, and cannot be further subdivided into smaller values.

It is noteworthy that, for filters with deeper observations, such as $g$, $r$, $J0660$, $i$, $J0861$, and $z$, brighter sources begin to saturate (see Figure \ref{fig:uss_gaia_predict}). The  $J0378$, $J0395$, $J0410$, $J0430$, and $J0515$ filters do not exhibit saturated objects within the same magnitude range.

While the photometric errors remain around 0.02 mag at magnitude 14 for the filters with deeper observations, a non-negligible photometric error is observed within the same magnitude range for shallower filters. Despite the fact that the redder filters reach a  suggested fainter limit for the use of USS DR1 photometry, we explore the data for stars with fainter magnitudes in the subsequent sections.

\subsection{Detection Completeness}

The detection completeness of the USS filters is determined for the Gaia DR3 $G$-band for bins of magnitudes. Figure \ref{fig:uss_completeness} shows the average completeness for the 12 bands in relation to the $G$-band in bins of magnitude. The completeness is around a hundred percent for all bins for the filters $g$, $r$, $J0515$, $J0660$, $i$, $J0861$, and $z$.  For filters $u$, $J0378$, $J0395$, $J0410$, and $J0430$ the detection completeness reach about 100\% until $G < 11$, and decreases significantly for fainter sources. 
For instance, for some filters, it is not possible to detect approximately 60\% of the sources at around $G \sim 14$.  This highlights a significant constraint on the use of USS DR1 data.

\begin{figure*}[ht!]
\centering
\includegraphics[width=\textwidth]{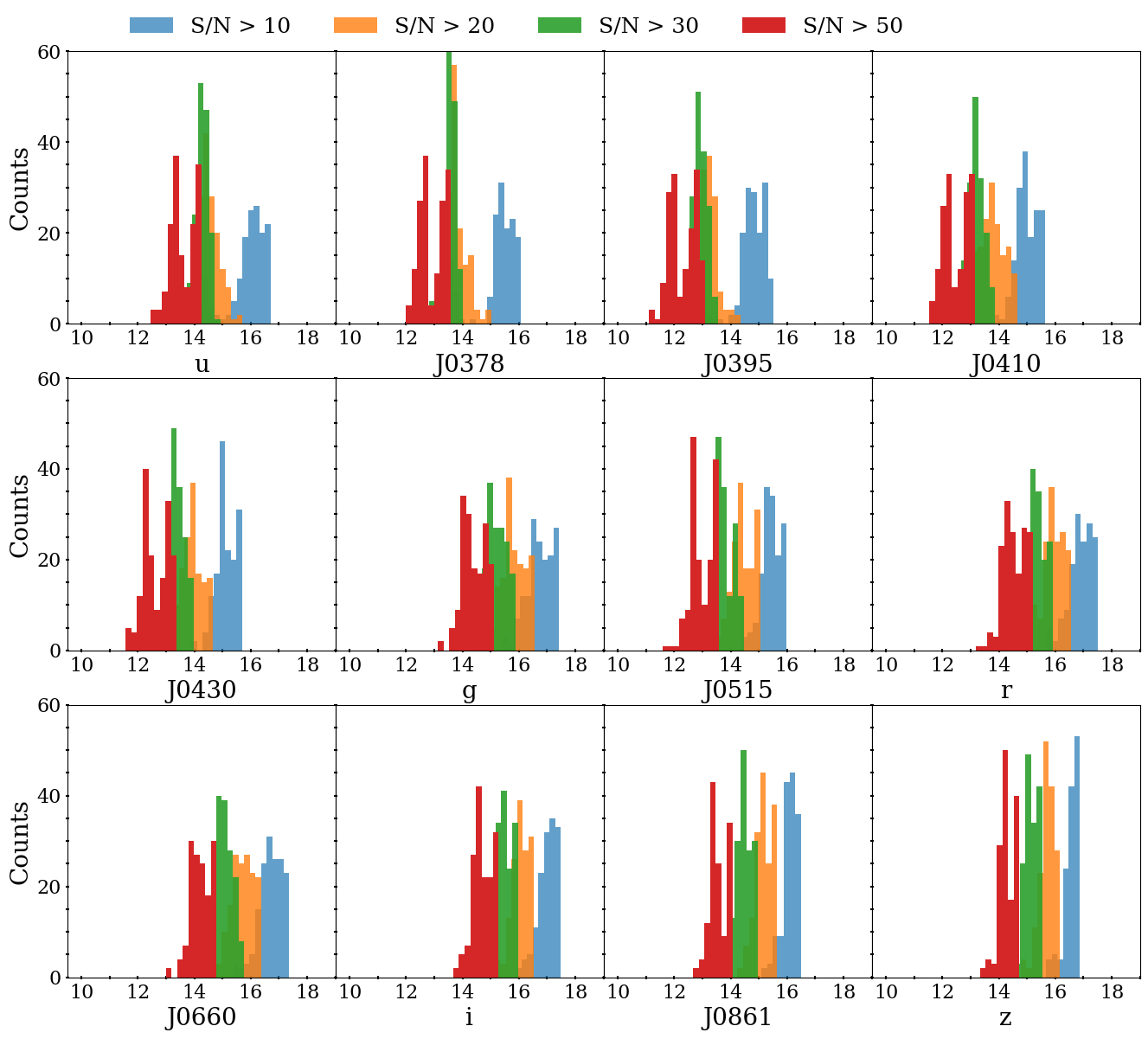}
\caption{Photometric depth in PStotal magnitudes for the 12 filters of the USS DR1 for the 
163 fields. In the histograms, distinct colors represent different S/N threshold values of 10, 20, 30, and 50, represented in blue, orange, green, and red, respectively.}
\label{fig:uss_depth}
\end{figure*}

\subsection{Depth}

The photometric depths were derived from the PStotal magnitudes of objects observed in the 163 fields from USS DR1. Although this provides a general overview of the depth, for the fainter magnitudes, the errors from the S-PLUS survey are lower than those provided by the USS DR1 due to the observations being taken in better photometric conditions and the co-adding of images. 

To obtain these measurements, the photometric depths for the 12 bands were calculated in each field, considering four maximum S/N values ($<$ 10, $<$ 20, $<$ 30, and $<$ 50). For each filter and S/N value, the peak of the magnitude distribution was estimated using a kernel density estimator, applied individually to all fields. Figure \ref{fig:uss_depth} shows the distribution of the estimated photometric depths. 

\subsection{Contents of USS DR1}
\label{sec:dr1}

The S-PLUS website\footnote{\href{https://splus.cloud}{https://splus.cloud}} serves as an interface to access the USS DR1 catalog, and provides a detailed description of the data, documentation of access tools, and example queries in ADQL for the database. Meta-information about object morphology, astrometry, and photometric measurements, along with relevant uncertainties, are presented in the documentation available on the website. It is important to note that the catalog shares the same columns as the single mode from the S-PLUS survey. For a comprehensive description of these columns, the reader may access to the documentation \footnote{\href{https://splus.cloud/documentation/uss}{https://splus.cloud/documentation/uss}}.

The first data release of the USS includes observations of 163 fields, covering a total area of $\sim$ 324 deg$^{2}$, across the 12 bands. The catalog contains approximately 1 million detections, from which $\sim$ 63,000 sources have $r < 14$.

\section{Search for Metal-Poor Stars}
\label{sec:search}

As outlined in the Introduction, EMP and UMP stars play a significant role in addressing various challenges within the fields of Stellar and Galactic Archaeology. The USS offers a valuable opportunity to identify relatively bright low-metallicity stars and conduct spectroscopic analyses to obtain chemical information that is challenging to access with fainter objects. This section describes ongoing efforts to find EMP and UMP candidates for future spectroscopic follow-up programs.

\subsection{Data Quality Control Cuts }
\label{sec:qualitycontrol}

We performed a series of quality-control cuts on the USS data to ensure the reliable selection of relatively bright EMP and UMP stars. Only sources with a high probability of being a star (\texttt{CLASS\_STAR\_R} $\geq 0.90$), not saturated ({\tt SEX\_FLAGS\_filter < 4}, for all filters), and with a precise magnitude estimation ($\texttt{e\_filter\_PStotal} \leq 0.2$) were selected. Additionally, we restricted the sample to specific color ranges ($0.2 \leq \texttt{g\_PStotal-i\_PStotal} \leq 1.4$; $0.3 \leq \texttt{J0410\_PStotal-J0861\_PStotal} \leq 3.5$, see \citealt{Placco2022}) to eliminate potential contamination from white dwarfs and A-type stars at the blue end, and objects cooler than $T_{\rm eff} \sim 4000$\,K \citep{Yanny2009} at the red end. After applying these criteria, our final sample comprises 45,520 stars. An ADQL query is provided in the Appendix \ref{app:query} to retrieve this sample through the S-PLUS website.

\subsection{Methodology and Target Selection}
\label{sec:method}

In our search for relatively bright EMP and UMP stars, we employ the metallicity- and temperature-sensitive filters available in the USS. In particular, the colors obtained using the narrow-band \texttt{J0395} filter have proven to be effective in separating different metallicity regimes. The validation of the Javalambre system for estimating stellar metallicity has recently been confirmed through different techniques \citep{Whitten2019,whitten2021,Galarza2022}. Moreover, the efficacy of detecting low-metallicity stars using narrow bands has been validated by employing color-color diagrams, which enable the identification of metal-poor stars \citep{Placco2022, Almeida-Fernandes2023}, leading to the discovery of EMP and UMP stars \citep{Placco2021, Placco2023}.

\begin{figure}
\centering
\includegraphics[width=\columnwidth]{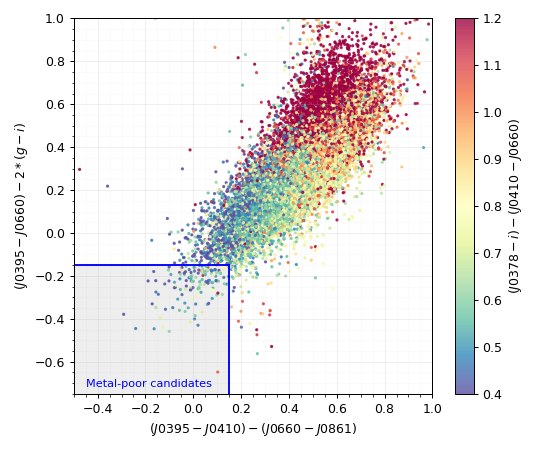}
\caption{Selection of low-metallicity stellar candidates in USS DR1, following the criteria proposed by \citet{Placco2022}: \texttt{(J0395-J0410)-(J0660-J0861)} $\leq 0.15$ and \texttt{(J0395-J0660)-2x(g-i)} $\leq -0.15$. The points are color-coded according to the \texttt{(J0378-i)-(J0410-J0660)} color; only stars with measurements below 0.8 in this color are included in our final selection.}
\label{fig:metalpoor_candidates}
\end{figure}

\cite{Placco2022}, following the work of \cite{Pristine1}, selected the color combination \texttt{(J0395-J0660)-2x(g-i)} to distinguish  stars in different metallicity regimes. Additionally, \cite{Placco2022} improved the color combination used by \cite{Pristine1} and \cite{Dacosta2019} by employing \texttt{(J0395-J0410)-(J0660-J0861)}, which increased the sensitivity to temperature variations. This refined color combination resulted in a success rate of approximately 83\% the identification of stars with $[\text{Fe/H}] < -2.0$. In Figure~\ref{fig:metalpoor_candidates}, we select candidates in the aforementioned color-color space by applying the same selection proposed by \citet{Placco2022}: \texttt{(J0395-J0410)-(J0660-J0861)} $\leq 0.15$ and \texttt{(J0395-J0660)-2x(g-i)} $\leq -0.15$. In this region, we identified a total of 152 stars, from which 91, 25, 7, and 2 targets have $r$-band magnitudes brighter than 14, 13, 12, and 11, respectively.

Additionally, \cite{Placco2022} observed that a substantial fraction of stars with $[\text{Fe/H}] > -1.0$ exhibit higher temperatures (i.e., $T_{\rm eff} > 5900$\,K). Taking this into account, \cite{Placco2022} identified a combination of temperature- and metallicity-sensitive filters that can further improve the success rate for identifying stars with $[\text{Fe/H}] < -2.0$ by removing stars with $T_{\rm eff} > 5900$\,K. Taking this into account, we restricted our target selection to retain only objects that have \texttt{(J0378-i)-(J0410-J0660)} $< 0.80$. This resulted in the selection of 140 stars, of which 91, 25, 7, and 2 targets have $r$-band magnitudes brighter than 14, 13, 12, and 11, respectively.

\citet{Placco2022} and \citet{Almeida-Fernandes2023} have shown that these color selections result in a purity of $\sim 16\%$ for EMP or UMP stars. Assuming a similar metallicity distribution for the magnitude range of the USS DR1, we can expect around 20 relatively bright stars ($r\leq14$) in our selection of EMP and UMP stars, which translates to 0.4 stars per square degree. The USS is expected to cover an area of 9300 deg$^2$ by the end of the survey, resulting in around 600 relatively bright EMP or UMP stars suitable for high-resolution spectroscopic follow-up.

To compare our selection with photometric metallicities, we opted to use the catalog of \citet{Andrae2023} instead of relying on metal-poor catalogs \citep{Matsuno2022, Viswanathan2023, Yao2024}. Andrae et al. catalog covers a wide range of metallicities and provides insight into the contamination of rich stars in the region. We cross-matched it with our more conservative selection of 140 stars, resulting in 112 stars in common. Within this subset, we observe that 60\% are VMP or EMP stars for $r < 13$, and this percentage drops to 50\% for $r < 14$. These numbers provide support for the success of our color-based selections, which will continue to provide candidates for spectroscopic follow-up.

\section{Conclusion}
\label{sec:conclusion}

This paper presents the first data release of the S-PLUS Short Survey (USS), which will cover approximately 9300 deg$^{2}$ of the Southern Hemisphere sky. The survey utilizes the Javalambre 12-band magnitude system, including both narrow and medium-band and broad-band filters. This data release contains the data from 163 observed fields, totaling $\sim$324 deg$^{2}$ and detecting a total of $\sim$ 1 million sources.

The photometric quality is ensured by a small average offset of 15 mmag between the USS observed magnitudes and those predicted by GaiaXPy, considering the weather conditions and exposure times. The uncertainties vary, ranging from a few mmag at the bright end to less than 0.02 and $\sim$ 0.04 mag at magnitudes 12 and 14, respectively. We advise caution in using objects brighter than $r = 10$, due to saturation, and objects fainter than $r = 14$, due to magnitude errors. In this faint regime, we recommend using data from the S-PLUS Main 
Survey for more reliable results.  We note that the use of the combination of the narrow and medium-band filter fluxes with Gaia BP/RP fluxes (rather than using the broad-band $ugriz$ filters, which saturate for stars brighter than $r \sim 10$), can extend our bright limit by several magnitudes, to $r \sim 7-8$. Further improvement may be obtained by fitting the wings of the PSFs, and extrapolating to account for the saturated flux, for even brighter stars. 

To identify relatively bright EMP and UMP stars within the USS DR1 data, we apply the color-magnitude combinations (\texttt{(J0395-J0660)-2x(g-i)} and \texttt{(J0395-J0410)-(J0660-J0861)}) validated by \cite{Placco2022}. This method initially flags 152 stars in USS DR1, with 91, 25, 7, and 2 brighter than $r = $14, 13, 12, and 11, respectively. A refined selection, incorporating \texttt{(J0378-i)-(J0410-J0660)}, reduces these numbers to 140 stars, with 84, 23, 6, and 1 brighter than $r = $ 14, 13, 12, and 11, respectively. As estimated by \cite{Placco2022} and \cite{Almeida-Fernandes2023}, these strict criteria ensure a purity of approximately 16\% for EMP or UMP stars. Extrapolating this to the USS magnitude range, we expect to identify around 20 relatively bright EMP or UMP stars ($r\leq14$) in DR1, and on the order of 600 stars for subsequent high-resolution spectroscopic follow-up upon the completion of the survey.

Future USS data releases are expected to provide data from thousands of fields. The early scientific exploration of these data will undoubtedly enhance the quality of future USS data releases, particularly with the incorporation of Gaia BP/RP spectra as part of the calibration process.  We note that such a procedure has already been carried out for J-PLUS DR3 (Huang et al., in prep.), and is being finalized for application to S-PLUS DR4.  In addition to the determination of stellar parameters ($T_{\rm eff}$, $\log g$, and [Fe/H]), these techniques provide elemental-abundance ratio estimates for [C/Fe] and [Mg/Fe].  Additional elemental-abundance estimates (e.g., [N/Fe] and [Ca/Fe]) will be added in the near future.

\begin{acknowledgements}

H.D.P. thanks FAPESP Proc. 2018/21250-9. The work of V.M.P. is supported by NOIRLab, which is managed by the Association of Universities for Research in Astronomy (AURA) under a cooperative agreement with the National Science Foundation. F.A.-F. acknowledges funding for this work from FAPESP (procs 2018/20977-2). F.R.H. acknowledges FAPESP for the financial support via grants 2018/21661-9 and 2021/11345-5. S.R. thanks partial financial support from FAPESP (procs. 2015/50374-0 and 2020/15245-2). T.C.B. acknowledges partial support from grant PHY 14-30152; Physics Frontier Center/JINA Center for the Evolution of the Elements (JINA-CEE), and from OISE-1927130: The International Research Network for Nuclear Astrophysics (IReNA), awarded by the US National Science Foundation. J.A. acknowledges funding from the European Research Council (ERC) under the European Union’s Horizon 2020 research and innovation programme (grant agreement No. 852839). G.L. acknowledges FAPESP (proc. 2021/10429-0).  A.W. acknowledges funding from the European Research Council (ERC) under the European Union's Horizon 2020 research and innovation program (grant agreement No. 833824, GASP project). L.BeS. acknowledges the support provided by the Heising Simons Foundation through the Barbara Pichardo Future Faculty Fellowship from grant \# 2022-3927. S.D. acknowledges CNPq/MCTI for grant 306859/2022-0. A.A.C. acknowledges financial support from the Severo Ochoa grant CEX2021-001131-S funded by MCIN/AEI/10.13039/501100011033. H.D.P. and R.S. acknowledge support from the National Science Centre, Poland, project 2019/34/E/ST9/00133.

The S-PLUS project, including the T80-South robotic telescope and the S-PLUS scientific survey was founded as a partnership between the Fundação de Amparo à Pesquisa do Estado de São Paulo (FAPESP), the Observatório Nacional (ON), the Federal University of Sergipe (UFS), and the Federal University of Santa Catarina (UFSC), with important financial and practical contributions from other collaborating institutes in Brazil, Chile (Universidad de La Serena), and Spain (Centro de Estudios de Física del Cosmos de Aragón, CEFCA). We further acknowledge financial support from the São Paulo Research Foundation (FAPESP), Fundação de Amparo à Pesquisa do Estado do RS (FAPERGS), the Brazilian National Research Council (CNPq), the Coordination for the Improvement of Higher Education Personnel (CAPES), the Carlos Chagas Filho Rio de Janeiro State Research Foundation (FAPERJ), and the Brazilian Innovation Agency (FINEP). Members of the S-PLUS collaboration are grateful for the contributions from CTIO staff in helping in the construction, commissioning, and maintenance of the T80-South telescope and camera.

\end{acknowledgements}

\bibliographystyle{aa}
\bibliography{bibliography.bib}

\appendix
\onecolumn
\section{S-PLUS Cloud Query}
\label{app:query}

The following ADQL query is used to download stars with photometry in all twelve bands of the USS:

\begin{verbatim}

SELECT 
    id, ra, dec,
    r_pstotal, g_pstotal, i_pstotal, z_pstotal, u_pstotal,
    j0378_pstotal, j0395_pstotal, j0410_pstotal,
    j0430_pstotal, j0515_pstotal, j0660_pstotal, j0861_pstotal
FROM 
    "usdr1"."usdr1"
WHERE 
    class_star_r >= 0.9 
    AND sex_flags_r < 4 
    AND sex_flags_g < 4 
    AND sex_flags_i < 4 
    AND sex_flags_u < 4 
    AND sex_flags_z < 4 
    AND sex_flags_j0378 < 4 
    AND sex_flags_j0395 < 4 
    AND sex_flags_j0410 < 4
    AND sex_flags_j0430 < 4 
    AND sex_flags_j0515 < 4 
    AND sex_flags_j0660 < 4 
    AND sex_flags_j0861 < 4
    AND e_r_pstotal <= 0.2 
    AND e_g_pstotal <= 0.2 
    AND e_i_pstotal <= 0.2 
    AND e_u_pstotal <= 0.2 
    AND e_z_pstotal <= 0.2 
    AND e_j0378_pstotal <= 0.2 
    AND e_j0395_pstotal <= 0.2 
    AND e_j0410_pstotal <= 0.2
    AND e_j0430_pstotal <= 0.2 
    AND e_j0515_pstotal <= 0.2 
    AND e_j0660_pstotal <= 0.2 
    AND e_j0861_pstotal <= 0.2
    AND (g_pstotal - i_pstotal) BETWEEN 0.2 AND 1.4
    AND (j0410_pstotal - j0861_pstotal) BETWEEN 0.3 AND 3.5
    AND r_pstotal != -99 AND g_pstotal != -99 AND i_pstotal != -99 
    AND z_pstotal != -99 AND u_pstotal != -99 AND j0378_pstotal != -99 
    AND j0395_pstotal != -99 AND j0410_pstotal != -99 AND j0430_pstotal != -99 
    AND j0515_pstotal != -99 AND j0660_pstotal != -99 AND j0861_pstotal != -99

\end{verbatim}

\section{Catalog of USS DR1 EMP/UMP candidates}
\label{app:exoplanets}

Table 4 presents a catalog comprising 140 EMP/UMP candidate stars. The table includes the following columns: ID, RA, DEC and \texttt{\{filter\}\_PStotal}.

\newpage

\clearpage


\renewcommand{\arraystretch}{1.0}
\setlength{\tabcolsep}{0.3em}

\begin{TableNotes}
\scriptsize
\item[] Notes: 
\item[] The magnitude columns represent values of PStotal and are not corrected for extinction.
\end{TableNotes}

\footnotesize
\onecolumn
\begin{landscape}
\begin{longtable}{>{\footnotesize}c >{\footnotesize}c >{\footnotesize}c >{\footnotesize}c >{\footnotesize}c >{\footnotesize}c >{\footnotesize}c >{\footnotesize}c >{\footnotesize}c >{\footnotesize}c >{\footnotesize}c >{\footnotesize}c >{\footnotesize}c >{\footnotesize}c >{\footnotesize}c}
\caption{Catalog of 140 EMP and UMP candidates} \\
\toprule
ID & RA                & DEC     & $r$ & $g$ & $i$ & $z$ & $u$ & $J0378$ & $J0395$ & $J0410$ & $J0430$ & $J0515$ & $J0660$ & $J0861$  \\
& (Deg) & (Deg) &  &  &  &  &  & & & & & & &             \\ \midrule \midrule
\endfirsthead
\multicolumn{15}{c}%
{{\bfseries Table \thetable\ }\textit{(continued)}} \\
\toprule
 ID & RA                & DEC     & $r$ & $g$ & $i$ & $z$ & $u$ & $J0378$ & $J0395$ & $J0410$ & $J0430$ & $J0515$ & $J0660$ & $J0861$  \\
 & (Deg) & (Deg) &  &  &  &  &  & & & & & & &               \\ \midrule \midrule
\endhead
\hline 
\insertTableNotes  \\
\multicolumn{15}{c}{{Continued on next page}}
\endfoot

\hline 
\insertTableNotes \\
\endlastfoot

	    SDR1\_SHORTS-STRIPE82\_0004\_0000006 & 1.6576471 & 1.3953081 & 13.12 & 13.59 & 12.89 & 12.81 & 14.55 & 14.2 & 14.18 & 13.92 & 13.8 & 13.44 & 13.05 & 12.87 \\ 
        SDR1\_SHORTS-STRIPE82\_0004\_0000108 &1.1896018 & 1.0214283 & 13.41 & 13.97 & 13.17 & 13.09 & 15.18 & 14.71 & 14.66 & 14.46 & 14.28 & 13.78 & 13.36 & 13.13 \\ 
        SDR1\_SHORTS-STRIPE82\_0007\_0000510 & 3.6443336 & -0.71293634 & 13.35 & 13.96 & 13.24 & 12.96 & 14.53 & 14.56 & 14.46 & 14.27 & 14.17 & 13.67 & 13.29 & 13.04 \\ 
        SDR1\_SHORTS-STRIPE82\_0008\_0000096 & 4.5058804 & 1.0855886 & 12.94 & 13.62 & 12.67 & 12.52 & 14.95 & 14.44 & 14.43 & 14.15 & 13.98 & 13.4 & 12.91 & 12.56 \\ 
        SDR1\_SHORTS-STRIPE82\_0008\_0000210 & 4.401828 & 0.15610152 & 13.92 & 14.56 & 13.67 & 13.55 & 15.89 & 15.54 & 15.16 & 14.97 & 14.92 & 14.33 & 13.83 & 13.61 \\ 
        SDR1\_SHORTS-STRIPE82\_0011\_0000366 & 7.6760125 & -1.0947679 & 13.54 & 14.14 & 13.35 & 13.27 & 15.13 & 14.75 & 14.61 & 14.41 & 14.35 & 13.9 & 13.49 & 13.3 \\ 
        SDR1\_SHORTS-STRIPE82\_0012\_0000087 & 7.4372125 & 1.1561842 & 12.63 & 13.34 & 12.37 & 12.25 & 15.23 & 14.33 & 14.07 & 14.12 & 13.87 & 13.15 & 12.56 & 12.29 \\ 
        SDR1\_SHORTS-STRIPE82\_0014\_0000254 & 7.853146 & 0.20201044 & 13.76 & 14.2 & 13.71 & 13.66 & 15.12 & 14.68 & 14.47 & 14.31 & 14.25 & 13.86 & 13.74 & 13.62 \\ 
        SDR1\_SHORTS-STRIPE82\_0018\_0000370 & 10.730438 & 0.23756407 & 12.14 & 12.61 & 11.96 & 11.91 & 13.79 & 13.35 & 13.2 & 12.97 & 12.85 & 12.45 & 12.1 & 11.93 \\ 
        SDR1\_SHORTS-STRIPE82\_0018\_0000459 & 11.594033 & 0.44198582 & 13.96 & 14.47 & 13.72 & 13.61 & 15.64 & 15.11 & 15.0 & 14.78 & 14.79 & 14.25 & 13.87 & 13.65 \\ 
        SDR1\_SHORTS-STRIPE82\_0018\_0000464 & 10.841239 & 0.4611773 & 14.47 & 14.82 & 14.3 & 14.35 & 15.9 & 15.69 & 15.23 & 15.15 & 15.12 & 14.7 & 14.48 & 14.35 \\ 
        SDR1\_SHORTS-STRIPE82\_0020\_0000462 & 13.056526 & 0.35977298 & 12.42 & 13.08 & 12.13 & 12.01 & 14.62 & 14.08 & 13.95 & 13.64 & 13.51 & 12.88 & 12.33 & 12.04 \\ 
        SDR1\_SHORTS-STRIPE82\_0021\_0000446 & 14.485071 & -1.0759861 & 14.69 & 15.11 & 14.6 & 14.6 & 16.0 & 15.78 & 15.55 & 15.4 & 15.27 & 14.97 & 14.73 & 14.58 \\ 
        SDR1\_SHORTS-STRIPE82\_0023\_0000551 & 16.080364 & -0.78245586 & 14.72 & 15.37 & 14.47 & 14.33 & 16.68 & 16.11 & 16.02 & 15.8 & 15.65 & 15.05 & 14.66 & 14.42 \\ 
        SDR1\_SHORTS-STRIPE82\_0025\_0000113 & 16.50135 & -0.30944324 & 14.32 & 14.63 & 14.22 & 14.2 & 15.84 & 15.18 & 14.89 & 14.76 & 14.66 & 14.54 & 14.33 & 14.26 \\ 
        SDR1\_SHORTS-STRIPE82\_0028\_0000295 & 19.048897 & 0.7529943 & 14.53 & 14.94 & 14.38 & 14.32 & 15.93 & 15.56 & 15.24 & 15.11 & 15.07 & 14.66 & 14.51 & 14.31 \\ 
        SDR1\_SHORTS-STRIPE82\_0031\_0000252 & 20.59981 & -0.6502333 & 13.53 & 14.03 & 13.42 & 13.33 & 14.91 & 14.74 & 14.46 & 14.26 & 14.24 & 13.82 & 13.52 & 13.34 \\ 
        SDR1\_SHORTS-STRIPE82\_0033\_0000095 & 21.968967 & -0.27128702 & 14.49 & 14.77 & 14.43 & 14.32 & 15.53 & 15.33 & 14.92 & 14.98 & 14.85 & 14.63 & 14.45 & 14.45 \\ 
        SDR1\_SHORTS-STRIPE82\_0034\_0000141 & 22.496855 & 0.10427119 & 13.45 & 14.01 & 13.18 & 13.05 & 15.22 & 14.78 & 14.71 & 14.33 & 14.27 & 13.74 & 13.38 & 13.1 \\ 
        SDR1\_SHORTS-STRIPE82\_0034\_0000149 & 22.547462 & 0.14666766 & 10.83 & 11.2 & 10.64 & 10.7 & 12.23 & 11.86 & 11.71 & 11.45 & 11.43 & 11.0 & 10.8 & 10.67 \\ 
        SDR1\_SHORTS-STRIPE82\_0050\_0000042 & 34.65409 & 1.3091509 & 14.62 & 15.05 & 14.58 & 14.57 & 15.9 & 15.41 & 15.37 & 15.22 & 15.26 & 14.85 & 14.69 & 14.58 \\ 
        SDR1\_SHORTS-STRIPE82\_0050\_0000320 & 33.967308 & 0.735502 & 14.17 & 14.56 & 14.02 & 14.03 & 15.4 & 15.06 & 14.96 & 14.77 & 14.78 & 14.37 & 14.16 & 14.01 \\ 
        SDR1\_SHORTS-STRIPE82\_0051\_0000249 & 35.983967 & -1.3795805 & 14.72 & 15.24 & 14.53 & 14.44 & 16.17 & 15.82 & 15.85 & 15.58 & 15.49 & 14.97 & 14.64 & 14.45 \\ 
        SDR1\_SHORTS-STRIPE82\_0056\_0000498 & 38.236706 & 0.33034867 & 14.25 & 14.93 & 13.94 & 13.78 & 16.58 & 15.92 & 15.84 & 15.58 & 15.44 & 14.73 & 14.16 & 13.83 \\ 
        SDR1\_SHORTS-STRIPE82\_0057\_0000234 & 39.839863 & -1.2754283 & 14.62 & 14.92 & 14.56 & 14.6 & 15.86 & 15.4 & 15.14 & 15.15 & 14.96 & 14.71 & 14.64 & 14.62 \\ 
        SDR1\_SHORTS-STRIPE82\_0057\_0000248 & 39.55139 & -1.2515044 & 14.47 & 14.9 & 14.36 & 14.36 & 15.82 & 15.59 & 15.36 & 15.19 & 15.15 & 14.67 & 14.47 & 14.32 \\ 
        SDR1\_SHORTS-STRIPE82\_0062\_0000438 & 43.19201 & 0.35812354 & 14.74 & 15.15 & 14.59 & 14.59 & 16.2 & 15.94 & 15.5 & 15.39 & 15.34 & 15.04 & 14.68 & 14.56 \\ 
        SDR1\_SHORTS-STRIPE82\_0072\_0000285 & 49.018974 & 0.06775388 & 13.77 & 14.38 & 13.59 & 13.47 & 15.59 & 15.07 & 14.96 & 14.65 & 14.55 & 14.13 & 13.73 & 13.51 \\ 
        SDR1\_SHORTS-STRIPE82\_0074\_0000206 & 50.69091 & 0.70975673 & 14.87 & 15.29 & 14.7 & 14.67 & 16.13 & 15.93 & 15.67 & 15.57 & 15.51 & 15.08 & 14.87 & 14.69 \\ 
        SDR1\_SHORTS-STRIPE82\_0079\_0000011 & 54.621475 & -0.024894878 & 14.5 & 14.93 & 14.35 & 14.34 & 15.82 & 15.56 & 15.39 & 15.23 & 15.06 & 14.85 & 14.49 & 14.33 \\ 
        SDR1\_SHORTS-STRIPE82\_0084\_0000084 & 58.519707 & 1.2353834 & 13.8 & 14.41 & 13.47 & 13.36 & 15.52 & 15.2 & 15.02 & 14.93 & 14.79 & 14.19 & 13.73 & 13.41 \\ 
        SDR1\_SHORTS-STRIPE82\_0086\_0000253 & 59.402622 & 0.06786838 & 14.73 & 15.43 & 14.35 & 14.21 & 16.95 & 16.48 & 16.12 & 16.08 & 15.81 & 15.27 & 14.59 & 14.26 \\ 
        SDR1\_SHORTS-STRIPE82\_0087\_0000039 & 300.19638 & -0.03680347 & 14.86 & 15.27 & 14.71 & 14.68 & 16.55 & 15.85 & 15.58 & 15.47 & 15.38 & 15.12 & 14.86 & 14.64 \\ 
        SDR1\_SHORTS-STRIPE82\_0087\_0000408 & 300.4243 & -0.32248077 & 14.47 & 14.91 & 14.28 & 14.18 & 15.98 & 15.74 & 15.3 & 15.27 & 15.22 & 14.79 & 14.45 & 14.23 \\ 
        SDR1\_SHORTS-STRIPE82\_0087\_0001390 & 301.11063 & -0.83181006 & 13.35 & 14.01 & 13.09 & 12.97 & 15.49 & 14.88 & 14.72 & 14.4 & 14.36 & 13.77 & 13.28 & 12.99 \\ 
        SDR1\_SHORTS-STRIPE82\_0088\_0000332 & 300.17416 & 1.2281477 & 14.23 & 14.92 & 13.93 & 13.77 & 16.61 & 16.04 & 15.7 & 15.52 & 15.33 & 14.67 & 14.16 & 13.82 \\ 
        SDR1\_SHORTS-STRIPE82\_0088\_0000652 & 299.81244 & 1.0351738 & 14.69 & 15.15 & 14.53 & 14.45 & 16.51 & 15.96 & 15.5 & 15.56 & 15.32 & 14.98 & 14.67 & 14.45 \\ 
        SDR1\_SHORTS-STRIPE82\_0088\_0001940 & 301.00873 & 0.5642095 & 14.36 & 14.92 & 14.15 & 14.03 & 16.26 & 15.75 & 15.33 & 15.32 & 15.18 & 14.68 & 14.3 & 14.03 \\ 
        SDR1\_SHORTS-STRIPE82\_0089\_0000203 & 301.38022 & -0.2050958 & 14.24 & 14.86 & 14.05 & 13.98 & 16.21 & 15.82 & 15.41 & 15.22 & 15.12 & 14.55 & 14.19 & 14.0 \\ 
        SDR1\_SHORTS-STRIPE82\_0089\_0000227 & 301.5744 & -0.22473826 & 13.31 & 13.97 & 13.02 & 12.85 & 15.2 & 14.8 & 14.68 & 14.4 & 14.32 & 13.75 & 13.23 & 12.91 \\ 
        SDR1\_SHORTS-STRIPE82\_0089\_0000849 & 301.89993 & -1.2271477 & 14.14 & 15.5 & 14.26 & 13.94 & 15.81 & 15.43 & 15.23 & 14.82 & 14.94 & 14.47 & 14.46 & 13.99 \\ 
        SDR1\_SHORTS-STRIPE82\_0090\_0000177 & 301.8522 & 1.2722871 & 13.7 & 14.38 & 13.4 & 13.27 & 15.9 & 15.44 & 15.25 & 14.93 & 14.83 & 14.13 & 13.66 & 13.3 \\ 
        SDR1\_SHORTS-STRIPE82\_0090\_0000417 & 302.25027 & 1.0935276 & 13.9 & 14.51 & 13.61 & 13.46 & 15.85 & 15.42 & 15.29 & 14.93 & 15.0 & 14.25 & 13.83 & 13.48 \\ 
        SDR1\_SHORTS-STRIPE82\_0091\_0000790 & 303.37555 & -0.62275493 & 13.07 & 13.92 & 12.86 & 12.94 & 15.12 & 14.55 & 14.31 & 14.03 & 13.92 & 13.38 & 13.3 & 12.77 \\ 
        SDR1\_SHORTS-STRIPE82\_0092\_0000166 & 303.1313 & 1.3386507 & 14.15 & 14.84 & 13.8 & 13.66 & 16.39 & 15.73 & 15.59 & 15.36 & 15.27 & 14.55 & 14.06 & 13.72 \\ 
        SDR1\_SHORTS-STRIPE82\_0092\_0000323 & 303.40805 & 1.2786375 & 13.95 & 14.63 & 13.64 & 13.45 & 16.13 & 15.45 & 15.48 & 15.2 & 15.01 & 14.4 & 13.88 & 13.53 \\ 
        SDR1\_SHORTS-STRIPE82\_0092\_0002329 & 303.2324 & 0.32604828 & 14.55 & 15.13 & 14.24 & 14.15 & 16.4 & 15.99 & 15.72 & 15.53 & 15.48 & 14.96 & 14.45 & 14.16 \\ 
        SDR1\_SHORTS-STRIPE82\_0092\_0003043 & 302.88132 & 0.6380521 & 14.17 & 14.76 & 13.89 & 13.78 & 16.22 & 15.79 & 15.48 & 15.15 & 15.08 & 14.6 & 14.1 & 13.79 \\ 
        SDR1\_SHORTS-STRIPE82\_0093\_0000912 & 305.3347 & -1.3431051 & 14.4 & 15.03 & 14.28 & 14.21 & 16.42 & 15.84 & 15.49 & 15.35 & 15.47 & 14.81 & 14.39 & 14.27 \\ 
        SDR1\_SHORTS-STRIPE82\_0093\_0001314 & 305.38776 & -0.5404583 & 14.35 & 14.82 & 14.21 & 14.14 & 15.81 & 15.5 & 15.21 & 15.1 & 15.1 & 14.61 & 14.35 & 14.16 \\ 
        SDR1\_SHORTS-STRIPE82\_0094\_0000946 & 304.33713 & 0.07345687 & 12.92 & 13.51 & 12.8 & 12.73 & 14.96 & 14.68 & 12.61 & 13.88 & 13.97 & 13.32 & 12.88 & 12.78 \\ 
        SDR1\_SHORTS-STRIPE82\_0095\_0000790 & 306.3834 & -1.2881083 & 14.23 & 14.92 & 13.98 & 13.84 & 16.22 & 15.98 & 15.64 & 15.41 & 15.32 & 14.7 & 14.19 & 13.87 \\ 
        SDR1\_SHORTS-STRIPE82\_0095\_0001623 & 306.72748 & -0.7906565 & 11.95 & 12.69 & 11.63 & 11.45 & 14.1 & 13.59 & 13.49 & 13.15 & 13.06 & 12.43 & 11.85 & 11.48 \\ 
        SDR1\_SHORTS-STRIPE82\_0096\_0000455 & 305.97064 & 1.074333 & 13.61 & 14.37 & 13.25 & 13.08 & 16.02 & 15.38 & 15.28 & 14.96 & 14.74 & 14.16 & 13.5 & 13.11 \\ 
        SDR1\_SHORTS-STRIPE82\_0098\_0000172 & 307.9158 & 1.2969278 & 14.65 & 15.26 & 14.45 & 14.34 & 16.45 & 16.26 & 15.75 & 15.56 & 15.62 & 15.07 & 14.62 & 14.34 \\ 
        SDR1\_SHORTS-STRIPE82\_0098\_0000565 & 308.04758 & 1.0347829 & 12.77 & 13.33 & 12.56 & 12.46 & 14.43 & 14.01 & 13.94 & 13.69 & 13.59 & 13.1 & 12.71 & 12.46 \\ 
        SDR1\_SHORTS-STRIPE82\_0098\_0000877 & 307.21344 & 0.013614895 & 14.54 & 15.14 & 14.28 & 14.19 & 16.53 & 16.17 & 15.74 & 15.44 & 15.47 & 15.04 & 14.45 & 14.22 \\ 
        SDR1\_SHORTS-STRIPE82\_0099\_0000111 & 309.29166 & -0.09919771 & 14.64 & 15.05 & 14.53 & 14.52 & 16.19 & 15.93 & 15.41 & 15.4 & 15.2 & 14.85 & 14.64 & 14.55 \\ 
        SDR1\_SHORTS-STRIPE82\_0100\_0001569 & 308.3723 & 0.4099422 & 14.23 & 14.8 & 14.05 & 13.91 & 16.01 & 15.69 & 15.36 & 15.16 & 15.16 & 14.62 & 14.21 & 13.94 \\ 
        SDR1\_SHORTS-STRIPE82\_0100\_0001740 & 308.47238 & 0.532722 & 14.46 & 14.92 & 14.34 & 14.31 & 16.17 & 15.86 & 13.47 & 15.15 & 15.15 & 14.75 & 14.42 & 14.32 \\ 
        SDR1\_SHORTS-STRIPE82\_0101\_0000043 & 310.42578 & -0.048552748 & 14.66 & 15.06 & 14.55 & 14.56 & 16.07 & 15.61 & 15.27 & 15.34 & 15.26 & 14.8 & 14.67 & 14.53 \\ 
        SDR1\_SHORTS-STRIPE82\_0101\_0000133 & 311.02887 & -0.15792824 & 13.64 & 14.3 & 13.43 & 13.28 & 15.54 & 15.18 & 15.05 & 14.68 & 14.58 & 13.99 & 13.6 & 13.3 \\ 
        SDR1\_SHORTS-STRIPE82\_0101\_0000211 & 310.84012 & -0.24697165 & 14.61 & 14.94 & 14.52 & 14.46 & 15.86 & 15.7 & 15.17 & 15.14 & 15.16 & 14.72 & 14.6 & 14.43 \\ 
        SDR1\_SHORTS-STRIPE82\_0101\_0000316 & 310.15854 & -0.37324297 & 14.1 & 14.62 & 13.89 & 13.79 & 15.66 & 15.37 & 15.18 & 14.92 & 14.91 & 14.39 & 14.05 & 13.78 \\ 
        SDR1\_SHORTS-STRIPE82\_0101\_0000484 & 310.93823 & -1.3851871 & 14.06 & 14.65 & 14.01 & 13.96 & 15.94 & 15.1 & 15.02 & 14.96 & 14.93 & 14.49 & 14.11 & 13.99 \\ 
        SDR1\_SHORTS-STRIPE82\_0101\_0001249 & 310.544 & -0.70014685 & 13.94 & 14.46 & 13.7 & 13.61 & 15.91 & 15.42 & 15.16 & 14.86 & 14.78 & 14.22 & 13.86 & 13.66 \\ 
        SDR1\_SHORTS-STRIPE82\_0102\_0000839 & 310.0794 & 0.16043985 & 14.07 & 14.7 & 13.84 & 13.74 & 15.81 & 15.57 & 15.4 & 15.1 & 15.1 & 14.52 & 14.0 & 13.77 \\ 
        SDR1\_SHORTS-STRIPE82\_0102\_0000916 & 310.59473 & 0.8105389 & 13.74 & 14.26 & 13.5 & 13.37 & 15.45 & 14.94 & 14.77 & 14.62 & 14.61 & 14.1 & 13.69 & 13.41 \\ 
        SDR1\_SHORTS-STRIPE82\_0103\_0000736 & 312.14856 & -1.2355337 & 13.46 & 14.08 & 13.2 & 13.08 & 15.59 & 15.13 & 14.84 & 14.62 & 14.43 & 13.84 & 13.37 & 13.12 \\ 
        SDR1\_SHORTS-STRIPE82\_0104\_0000498 & 312.22162 & 0.9273911 & 11.81 & 12.64 & 11.46 & 11.27 & 14.56 & 13.9 & 13.76 & 13.34 & 13.16 & 12.39 & 11.7 & 11.31 \\ 
        SDR1\_SHORTS-STRIPE82\_0104\_0001129 & 311.59122 & 0.3266088 & 14.83 & 15.2 & 14.67 & 14.64 & 16.18 & 15.85 & 15.53 & 15.46 & 15.43 & 15.0 & 14.79 & 14.63 \\ 
        SDR1\_SHORTS-STRIPE82\_0104\_0001367 & 311.8505 & 0.57589555 & 14.64 & 15.06 & 14.5 & 14.42 & 16.42 & 15.7 & 15.29 & 15.24 & 15.25 & 14.87 & 14.62 & 14.44 \\ 
        SDR1\_SHORTS-STRIPE82\_0104\_0001409 & 312.2212 & 0.6251173 & 14.62 & 15.17 & 14.45 & 14.39 & 16.52 & 15.93 & 15.47 & 15.55 & 15.44 & 14.98 & 14.6 & 14.39 \\ 
        SDR1\_SHORTS-STRIPE82\_0104\_0001476 & 312.32614 & 0.68522644 & 13.43 & 14.12 & 13.17 & 13.02 & 15.45 & 14.91 & 14.9 & 14.52 & 14.44 & 13.91 & 13.35 & 13.05 \\ 
        SDR1\_SHORTS-STRIPE82\_0105\_0000377 & 313.18854 & -0.38996223 & 13.95 & 14.47 & 13.73 & 13.63 & 15.54 & 15.3 & 15.03 & 14.78 & 14.81 & 14.29 & 13.89 & 13.63 \\ 
        SDR1\_SHORTS-STRIPE82\_0105\_0000483 & 313.69315 & -0.49006832 & 13.77 & 14.49 & 13.46 & 13.29 & 16.09 & 15.57 & 15.39 & 15.13 & 15.06 & 14.24 & 13.67 & 13.31 \\ 
        SDR1\_SHORTS-STRIPE82\_0105\_0000549 & 313.00925 & -1.3846456 & 14.9 & 15.29 & 14.78 & 14.76 & 16.49 & 16.19 & 15.63 & 15.56 & 15.6 & 15.05 & 14.89 & 14.78 \\ 
        SDR1\_SHORTS-STRIPE82\_0105\_0000658 & 312.69165 & -1.2699544 & 13.94 & 14.34 & 13.81 & 13.79 & 15.1 & 14.82 & 14.6 & 14.57 & 14.43 & 14.1 & 13.89 & 13.8 \\ 
        SDR1\_SHORTS-STRIPE82\_0106\_0000610 & 313.1019 & 0.18768471 & 13.26 & 13.94 & 12.98 & 12.8 & 15.32 & 14.79 & 14.51 & 14.42 & 14.25 & 13.72 & 13.17 & 12.86 \\ 
        SDR1\_SHORTS-STRIPE82\_0106\_0000712 & 313.80792 & 0.3431909 & 14.04 & 14.68 & 13.77 & 13.63 & 15.99 & 15.49 & 15.43 & 15.13 & 15.08 & 14.47 & 13.98 & 13.67 \\ 
        SDR1\_SHORTS-STRIPE82\_0106\_0000834 & 313.8282 & 0.52028793 & 13.52 & 14.18 & 13.31 & 13.2 & 15.55 & 15.24 & 14.9 & 14.61 & 14.56 & 13.91 & 13.46 & 13.24 \\ 
        SDR1\_SHORTS-STRIPE82\_0107\_0000040 & 314.83517 & -0.038442463 & 14.32 & 14.69 & 14.19 & 14.11 & 15.71 & 15.24 & 14.86 & 14.89 & 14.84 & 14.55 & 14.33 & 14.19 \\ 
        SDR1\_SHORTS-STRIPE82\_0107\_0000253 & 313.98013 & -0.3653406 & 14.69 & 15.04 & 14.58 & 14.62 & 16.11 & 15.86 & 15.33 & 15.29 & 15.2 & 14.91 & 14.69 & 14.64 \\ 
        SDR1\_SHORTS-STRIPE82\_0107\_0000478 & 315.06863 & -1.310828 & 14.44 & 14.87 & 14.3 & 14.26 & 15.92 & 15.48 & 15.13 & 15.02 & 15.0 & 14.63 & 14.43 & 14.32 \\ 
        SDR1\_SHORTS-STRIPE82\_0107\_0000942 & 314.30426 & -0.8858297 & 14.42 & 14.8 & 14.26 & 14.2 & 16.16 & 15.5 & 15.13 & 15.08 & 15.03 & 14.63 & 14.38 & 14.23 \\ 
        SDR1\_SHORTS-STRIPE82\_0108\_0000333 & 314.37265 & 1.144085 & 14.89 & 15.3 & 14.78 & 14.74 & 16.37 & 16.11 & 15.55 & 15.59 & 15.44 & 15.23 & 14.85 & 14.76 \\ 
        SDR1\_SHORTS-STRIPE82\_0108\_0000785 & 314.62006 & 0.07866466 & 12.34 & 13.1 & 12.0 & 11.86 & 14.8 & 14.29 & 14.13 & 13.69 & 13.56 & 12.84 & 12.24 & 11.89 \\ 
        SDR1\_SHORTS-STRIPE82\_0108\_0000985 & 314.34805 & 0.7968035 & 14.39 & 14.77 & 14.27 & 14.24 & 15.61 & 15.44 & 15.12 & 15.1 & 14.98 & 14.65 & 14.37 & 14.26 \\ 
        SDR1\_SHORTS-STRIPE82\_0108\_0001162 & 314.10263 & 0.2225285 & 12.31 & 14.89 & 14.01 & 13.94 & 13.96 & 13.75 & 13.58 & 13.14 & 15.13 & 14.56 & 14.12 & 12.13 \\ 
        SDR1\_SHORTS-STRIPE82\_0108\_0001229 & 314.0349 & 0.27821594 & 14.26 & 14.68 & 14.12 & 14.08 & 15.52 & 15.32 & 15.04 & 15.01 & 14.9 & 14.51 & 14.24 & 14.07 \\ 
        SDR1\_SHORTS-STRIPE82\_0109\_0000167 & 316.07025 & -0.13544065 & 14.38 & 15.0 & 14.18 & 14.06 & 16.3 & 16.02 & 15.46 & 15.4 & 15.22 & 14.74 & 14.33 & 14.08 \\ 
        SDR1\_SHORTS-STRIPE82\_0109\_0000326 & 315.71826 & -0.28916782 & 14.01 & 14.51 & 13.84 & 13.81 & 15.85 & 15.65 & 12.52 & 14.96 & 14.92 & 14.32 & 13.96 & 13.82 \\ 
        SDR1\_SHORTS-STRIPE82\_0109\_0000535 & 316.51276 & -0.50278807 & 14.06 & 14.62 & 13.83 & 13.77 & 15.65 & 15.36 & 15.06 & 14.88 & 14.9 & 14.41 & 13.99 & 13.76 \\ 
        SDR1\_SHORTS-STRIPE82\_0109\_0000726 & 315.98138 & -1.2344275 & 13.02 & 13.41 & 12.83 & 12.81 & 14.31 & 13.96 & 13.8 & 13.61 & 13.57 & 13.24 & 12.96 & 12.82 \\ 
        SDR1\_SHORTS-STRIPE82\_0109\_0000769 & 316.11676 & -0.8261581 & 13.95 & 14.58 & 13.67 & 13.53 & 15.86 & 15.42 & 15.14 & 14.97 & 14.9 & 14.32 & 13.88 & 13.56 \\ 
        SDR1\_SHORTS-STRIPE82\_0109\_0001384 & 316.65747 & -0.77296907 & 14.02 & 14.61 & 13.74 & 13.69 & 15.85 & 15.46 & 15.26 & 14.97 & 14.9 & 14.33 & 13.91 & 13.68 \\ 
        SDR1\_SHORTS-STRIPE82\_0109\_0001385 & 315.98273 & -0.7726212 & 13.86 & 14.36 & 13.68 & 14.42 & 15.68 & 15.34 & 15.4 & 15.44 & 14.71 & 14.18 & 14.57 & 14.45 \\ 
        SDR1\_SHORTS-STRIPE82\_0109\_0001443 & 315.9575 & -0.71563333 & 13.78 & 14.52 & 13.48 & 13.33 & 16.17 & 15.69 & 15.39 & 15.11 & 14.93 & 14.26 & 13.69 & 13.39 \\ 
        SDR1\_SHORTS-STRIPE82\_0113\_0001171 & 319.38544 & -0.9063646 & 13.3 & 14.06 & 12.99 & 12.84 & 15.8 & 15.29 & 15.1 & 14.68 & 14.59 & 13.8 & 13.2 & 12.87 \\ 
        SDR1\_SHORTS-STRIPE82\_0115\_0000110 & 320.79617 & -0.10950566 & 14.43 & 14.82 & 14.26 & 14.24 & 16.01 & 15.49 & 15.23 & 15.01 & 14.94 & 14.59 & 14.42 & 14.29 \\ 
        SDR1\_SHORTS-STRIPE82\_0116\_0000212 & 320.9774 & 1.0159353 & 14.62 & 14.95 & 14.49 & 14.44 & 15.87 & 15.45 & 15.23 & 15.28 & 15.04 & 14.73 & 14.6 & 14.47 \\ 
        SDR1\_SHORTS-STRIPE82\_0116\_0000705 & 320.87167 & 0.46007103 & 14.38 & 14.78 & 14.27 & 14.21 & 15.88 & 15.48 & 15.06 & 15.18 & 14.98 & 14.59 & 14.34 & 14.24 \\ 
        SDR1\_SHORTS-STRIPE82\_0116\_0000832 & 320.78775 & 0.700775 & 13.44 & 14.16 & 13.15 & 12.99 & 15.75 & 15.25 & 15.04 & 14.74 & 14.72 & 13.88 & 13.34 & 13.03 \\ 
        SDR1\_SHORTS-STRIPE82\_0118\_0000130 & 321.63882 & 1.2726451 & 13.81 & 14.48 & 13.55 & 13.4 & 15.93 & 15.32 & 15.3 & 14.91 & 14.81 & 14.22 & 13.73 & 13.39 \\ 
        SDR1\_SHORTS-STRIPE82\_0119\_0000063 & 323.6143 & -0.1691683 & 12.32 & 13.0 & 12.09 & 11.96 & 14.34 & 13.91 & 13.71 & 13.39 & 13.35 & 12.76 & 12.3 & 12.0 \\ 
        SDR1\_SHORTS-STRIPE82\_0119\_0000096 & 323.1112 & -0.29941627 & 14.06 & 14.54 & 13.96 & 13.98 & 15.59 & 15.38 & 14.92 & 14.8 & 14.71 & 14.23 & 14.08 & 13.91 \\ 
        SDR1\_SHORTS-STRIPE82\_0119\_0000280 & 323.64536 & -1.02573 & 14.24 & 14.66 & 14.2 & 14.17 & 15.52 & 15.11 & 14.79 & 14.72 & 14.8 & 14.48 & 14.26 & 14.22 \\ 
        SDR1\_SHORTS-STRIPE82\_0119\_0000315 & 323.36276 & -0.8233687 & 11.94 & 12.77 & 11.81 & 11.68 & 13.91 & 13.66 & 13.4 & 13.28 & 13.11 & 12.43 & 11.81 & 11.65 \\ 
        SDR1\_SHORTS-STRIPE82\_0120\_0000668 & 323.40384 & 0.3954084 & 12.03 & 12.35 & 11.93 & 11.9 & 13.17 & 12.86 & 12.61 & 12.54 & 12.44 & 12.2 & 12.0 & 11.93 \\ 
        SDR1\_SHORTS-STRIPE82\_0125\_0000766 & 327.80646 & -0.7942097 & 13.78 & 14.35 & 13.51 & 13.38 & 15.49 & 15.09 & 14.93 & 14.78 & 14.65 & 14.13 & 13.7 & 13.43 \\ 
        SDR1\_SHORTS-STRIPE82\_0128\_0000568 & 329.0807 & 0.19897743 & 13.55 & 13.95 & 13.44 & 13.41 & 14.8 & 14.51 & 14.29 & 14.13 & 14.12 & 13.81 & 13.54 & 13.41 \\ 
        SDR1\_SHORTS-STRIPE82\_0129\_0000049 & 330.816 & -0.0885891 & 14.05 & 14.64 & 13.85 & 13.74 & 15.9 & 15.47 & 15.32 & 15.1 & 14.95 & 14.5 & 14.03 & 13.76 \\ 
        SDR1\_SHORTS-STRIPE82\_0131\_0000580 & 331.396 & -0.7343162 & 13.7 & 14.38 & 13.4 & 13.21 & 16.04 & 15.38 & 15.25 & 14.93 & 14.79 & 14.15 & 13.6 & 13.27 \\ 
        SDR1\_SHORTS-STRIPE82\_0132\_0000445 & 332.25662 & 0.8650406 & 14.39 & 14.76 & 14.25 & 14.26 & 15.96 & 15.61 & 15.12 & 15.09 & 14.99 & 14.73 & 14.35 & 14.29 \\ 
        SDR1\_SHORTS-STRIPE82\_0132\_0000719 & 332.13974 & 0.710427 & 14.4 & 14.88 & 14.22 & 14.19 & 15.91 & 15.75 & 15.39 & 15.28 & 15.28 & 14.79 & 14.35 & 14.22 \\ 
        SDR1\_SHORTS-STRIPE82\_0135\_0000179 & 334.06198 & -0.4520497 & 14.08 & 14.6 & 13.86 & 13.81 & 15.94 & 15.38 & 15.22 & 15.19 & 14.88 & 14.41 & 13.99 & 13.82 \\ 
        SDR1\_SHORTS-STRIPE82\_0136\_0000079 & 333.86755 & 1.2393587 & 14.91 & 15.32 & 14.86 & 14.79 & 16.11 & 16.07 & 15.56 & 15.45 & 15.43 & 15.21 & 14.93 & 14.85 \\ 
        SDR1\_SHORTS-STRIPE82\_0136\_0000384 & 334.59164 & 0.7780391 & 14.31 & 14.98 & 14.0 & 13.85 & 16.71 & 16.08 & 15.75 & 15.6 & 15.36 & 14.81 & 14.21 & 13.87 \\ 
        SDR1\_SHORTS-STRIPE82\_0139\_0000295 & 337.55377 & -0.88559717 & 11.66 & 12.44 & 11.31 & 11.15 & 14.25 & 13.59 & 13.48 & 13.12 & 12.95 & 12.22 & 11.57 & 11.2 \\ 
        SDR1\_SHORTS-STRIPE82\_0141\_0000013 & 338.9892 & -0.030445319 & 14.66 & 15.06 & 14.57 & 14.51 & 16.28 & 15.6 & 15.31 & 15.25 & 15.27 & 14.84 & 14.65 & 14.48 \\ 
        SDR1\_SHORTS-STRIPE82\_0141\_0000105 & 339.17032 & -0.2691278 & 13.21 & 14.04 & 12.83 & 12.67 & 15.97 & 15.44 & 15.19 & 14.91 & 14.58 & 13.75 & 13.09 & 12.7 \\ 
        SDR1\_SHORTS-STRIPE82\_0143\_0000035 & 339.82462 & -0.119511425 & 13.91 & 14.55 & 13.66 & 13.53 & 15.81 & 15.43 & 15.28 & 15.14 & 14.95 & 14.26 & 13.85 & 13.54 \\ 
        SDR1\_SHORTS-STRIPE82\_0143\_0000283 & 340.15884 & -1.2710394 & 12.22 & 13.06 & 11.86 & 11.69 & 14.96 & 14.33 & 14.17 & 13.75 & 13.59 & 12.8 & 12.11 & 11.74 \\ 
        SDR1\_SHORTS-STRIPE82\_0147\_0000322 & 342.71426 & -1.0235988 & 14.7 & 15.07 & 14.53 & 14.48 & 15.96 & 15.79 & 15.49 & 15.31 & 15.25 & 14.92 & 14.68 & 14.5 \\ 
        SDR1\_SHORTS-STRIPE82\_0151\_0000121 & 346.32645 & -0.30182838 & 13.98 & 14.37 & 13.84 & 13.86 & 15.58 & 15.15 & 14.79 & 14.7 & 14.56 & 14.24 & 13.95 & 13.84 \\ 
        SDR1\_SHORTS-STRIPE82\_0152\_0000423 & 345.2313 & 0.21967787 & 12.38 & 12.8 & 12.2 & 12.23 & 13.74 & 13.4 & 13.27 & 13.08 & 13.02 & 12.68 & 12.32 & 12.18 \\ 
        SDR1\_SHORTS-STRIPE82\_0153\_0000034 & 346.71744 & -0.13730732 & 12.64 & 13.21 & 12.46 & 12.42 & 14.28 & 13.91 & 13.71 & 13.51 & 13.51 & 12.93 & 12.59 & 12.43 \\ 
        SDR1\_SHORTS-STRIPE82\_0153\_0000307 & 347.2783 & -0.8693464 & 14.01 & 14.33 & 13.82 & 13.8 & 15.58 & 15.15 & 14.7 & 14.67 & 14.62 & 14.16 & 13.94 & 13.88 \\ 
        SDR1\_SHORTS-STRIPE82\_0154\_0000195 & 347.83527 & 0.9204981 & 14.14 & 14.58 & 13.9 & 13.86 & 15.89 & 15.53 & 15.2 & 14.98 & 14.85 & 14.43 & 14.06 & 13.86 \\ 
        SDR1\_SHORTS-STRIPE82\_0157\_0000444 & 350.66132 & -0.9999145 & 14.18 & 14.52 & 14.07 & 14.04 & 15.44 & 15.06 & 14.83 & 14.77 & 14.68 & 14.4 & 14.18 & 14.06 \\ 
        SDR1\_SHORTS-STRIPE82\_0160\_0000250 & 351.24118 & 0.149503 & 14.12 & 14.61 & 13.98 & 14.05 & 15.58 & 15.32 & 15.1 & 15.0 & 14.77 & 14.47 & 14.12 & 14.04 \\ 
        SDR1\_SHORTS-STRIPE82\_0160\_0000271 & 351.37036 & 0.5066438 & 13.48 & 14.21 & 13.06 & 12.99 & 16.15 & 15.46 & 15.34 & 14.95 & 14.83 & 14.0 & 13.36 & 13.03 \\ 
        SDR1\_SHORTS-STRIPE82\_0161\_0000253 & 353.3934 & -1.3361326 & 12.97 & 13.43 & 12.88 & 12.83 & 14.23 & 13.98 & 13.81 & 13.59 & 13.56 & 13.21 & 12.96 & 12.85 \\ 
        SDR1\_SHORTS-STRIPE82\_0161\_0000324 & 353.44833 & -0.5359643 & 14.34 & 14.77 & 14.26 & 14.21 & 15.75 & 15.42 & 15.08 & 14.92 & 14.87 & 14.58 & 14.37 & 14.21 \\ 
        SDR1\_SHORTS-STRIPE82\_0161\_0000438 & 353.4584 & -0.9787993 & 14.73 & 15.18 & 14.66 & 14.6 & 16.23 & 15.75 & 15.58 & 15.4 & 15.26 & 15.08 & 14.78 & 14.58 \\ 
        SDR1\_SHORTS-STRIPE82\_0162\_0000159 & 353.3701 & 0.9590613 & 14.0 & 14.42 & 13.85 & 13.78 & 15.38 & 15.13 & 14.87 & 14.7 & 14.73 & 14.21 & 13.97 & 13.83 \\ 
        SDR1\_SHORTS-STRIPE82\_0163\_0000365 & 353.72775 & -0.95116305 & 12.64 & 13.21 & 12.46 & 12.37 & 14.34 & 13.91 & 13.87 & 13.57 & 13.48 & 12.99 & 12.6 & 12.37 \\ 
        SDR1\_SHORTS-STRIPE82\_0166\_0000308 & 355.3462 & 0.28945935 & 13.84 & 14.44 & 13.57 & 13.44 & 15.84 & 15.23 & 15.19 & 15.03 & 14.78 & 14.3 & 13.75 & 13.5 \\ 
        SDR1\_SHORTS-STRIPE82\_0166\_0000317 & 355.2404 & 0.34346882 & 13.44 & 13.93 & 13.27 & 13.21 & 15.22 & 14.94 & 13.01 & 14.45 & 14.3 & 13.78 & 13.36 & 13.26 \\ 
        SDR1\_SHORTS-STRIPE82\_0169\_0000403 & 358.82654 & -0.8964953 & 13.82 & 14.47 & 13.61 & 13.49 & 15.83 & 15.33 & 15.28 & 14.99 & 14.82 & 14.25 & 13.76 & 13.53 \\ 

\bottomrule

\end{longtable}
 \end{landscape}
\twocolumn



\end{document}